\documentclass[prb,
showpacs,
twocolumn,
floats,
10pt,
aps,
citeautoscript,
longbibliography,
floatfix]{revtex4-2}

\usepackage{placeins}
\usepackage{lipsum}
\usepackage{xcolor}
\usepackage[normalem]{ulem}
\usepackage{comment}
\usepackage{tabularx}
\usepackage{graphicx}
\usepackage{dcolumn}
\usepackage{bm}
\usepackage{booktabs}

\usepackage{bbm}
\usepackage{blindtext}
\usepackage{graphics}
\usepackage{verbatim}   
\usepackage{amsfonts}
\usepackage{amsmath}
\usepackage{amssymb}
\usepackage{mathtools}
\usepackage{adjustbox}
\usepackage{microtype} 
\allowdisplaybreaks 
\usepackage{xspace} 
\usepackage{multirow} 
\usepackage{tabstackengine}
\usepackage{tikz}
\setstackEOL{\cr}

\newcommand{\Eq}[1]{Eq.~\eqref{#1}}

\newcommand{\Fig}[1]{Fig.~\ref{#1}}

\newcommand{\Sec}[1]{Sec.~\ref{#1}}
\newcommand{\App}[1]{App.~\ref{#1}}

\def\md{\mathrm{d}}   
\def\mi{\mathrm{i}}   
\def\me{\mathrm{e}}   

\newcommand*{\ndots}{\kern-0.075em.\kern-0.05em.\kern-0.05em.}  
\newcommand*{\nidots}{.\kern-0.05em.\kern-0.05em.} 
\newcommand*{\ncdots}{\kern-0.15em\cdot\kern-0.2em\cdot\kern-0.2em\cdot\kern-0.15em}   

\NewDocumentCommand{\doubleI}{O{}}{\mathbbm{1}_{#1}}
\NewDocumentCommand{\doubleIb}{O{}}{{\overline{\mathbbm{1}}_{#1}}}
\NewDocumentCommand{\doubleIk}{O{}}{\mathbbm{1}^\ks_{\! #1}}
\NewDocumentCommand{\doubleId}{O{}}{\mathbbm{1}^\ds_{\! #1}}
\NewDocumentCommand{\doubleIp}{O{}}{\mathbbm{1}^\ps_{\! #1}}
\NewDocumentCommand{\doubleV}{O{}}{\mathbbm{V}_{\! #1}}
\NewDocumentCommand{\doubleVk}{O{}}{\mathbbm{V}^\ks_{\! #1}}
\NewDocumentCommand{\doubleVd}{O{}}{\mathbbm{V}^\ds_{\! #1}}
\NewDocumentCommand{\doubleVp}{O{}}{\mathbbm{V}^\ps_{\! #1}}
\NewDocumentCommand{\doublev}{o}{{\mathbbm{v}_{#1}}}
\NewDocumentCommand{\doubleVb}{o}{{\overline{\mathbbm{V}}_{\! #1}}}
\NewDocumentCommand{\doubleVt}{o}{{\widetilde{\mathbbm{V}}_{\! #1}}}
\NewDocumentCommand{\doubleVh}{o}{\widehat{{\mathbbm{V}}_{\! #1}}}
\NewDocumentCommand{\doubleW}{o}{\mathbbm{W}_{\! #1}}
\NewDocumentCommand{\doubleWk}{o}{\mathbbm{W}^\ks_{\! #1}}
\NewDocumentCommand{\doubleWd}{o}{\mathbbm{W}^\ds_{\! #1}}
\NewDocumentCommand{\doubleWb}{o}{{\overline{\mathbbm{W}}_{\! #1}}}
\NewDocumentCommand{\doubleWt}{o}{{\widetilde{\mathbbm{V}}_{\! #1}}}
\NewDocumentCommand{\doubleWh}{o}{{\widehat{\mathbbm{V}}_{\! #1}}}

\newcommand{\onedgrid}{W}
\newcommand{\bomega}{\boldsymbol{\omega}}
\newcommand{\bepsilon}{\boldsymbol{\epsilon}}
\newcommand{\bsigma}{\boldsymbol{\sigma}}
\newcommand{\omsum}[1]{\omega_{\overline{1}\ldots\overline{#1}}}
\newcommand{\kelsum}[1]{k_{\overline{1}\ldots\overline{#1}}}
\newcommand{\specfun}{S[\boldsymbol{\mathcal{O}}]}
\newcommand{\pspecfun}{S[\boldsymbol{\mathcal{O}}_p]}
\newcommand{\fullcorr}{G(\mi\bomega)}
\newcommand{\fullcorrK}{G^{\boldsymbol{k}}(\bomega)}
\newcommand{\gamcore}{\Gamma_{\mathrm{core}}}
\newcommand{\gambare}{\Gamma_0}
\newcommand{\opstring}{\boldsymbol{\mathcal{O}}}
\newcommand{\omint}[1]{{\omega}_{\bar{1}\cdots\bar{#1}}}
\newcommand{\epsint}[1]{{\epsilon}_{\bar{1}\cdots\bar{#1}}}
\newcommand{\Omint}[1]{{\Omega}_{\bar{1}\cdots\bar{#1}}}

\newcommand{\bomintthree}{{\omega}_{\bar{1}},{\omega}_{\bar{1}\bar{2}},{\omega}_{\bar{1}\bar{2}\bar{3}}}
\newcommand{\bepsintthree}{{\epsilon}_{\bar{1}},{\epsilon}_{\bar{1}\bar{2}},{\epsilon}_{\bar{1}\bar{2}\bar{3}}}
\newcommand{\epsinttwo}{{\epsilon}_{\bar{1}\bar{2}}}
\newcommand{\epsintthree}{{\epsilon}_{\bar{1}\bar{2}\bar{3}}}

\newcommand{\bretkernel}[1]{K_b^{[#1]}}
\newcommand{\retgp}{G^{[\lambda]}_p}
\newcommand{\imagshift}[1][i]{\gamma^{\lambda}_{0,#1}}
\newcommand{\kernelmat}{k_{\omint{i}\epsint{i}}}
\newcommand{\isokernel}{U_{\omega_i a_i}}
\newcommand{\sigmacut}{S_{\mathrm{cut}}}
\newcommand{\upup}{\uparrow\uparrow}
\newcommand{\updown}{\uparrow\downarrow}
\newcommand{\threetimes}[1]{#1\!\times\!#1\!\times\!#1}
\newcommand{\qbit}{\sigma_{\ell}}
\newcommand{\tcitol}{\tau}
\newcommand{\tcitoltxt}{\tau=}
\newcommand{\kclass}{\mathcal{K}}

\newcommand{\LMUMunich}{Arnold Sommerfeld Center for Theoretical Physics, Center for NanoScience, and Munich Center for Quantum Science and Technology, Ludwig-Maximilians-Universit\"at M\"unchen, 80333 Munich, Germany}

\newcommand{\mycomment}[1]{}
\newcommand{\memmb}[1]{\mbox{#1\,MB}}
\newcommand{\error}{\varepsilon}

\definecolor{darkgreen}{rgb}{0,0.5,0}
\definecolor{purple}{rgb}{0.6,0,0.5}
\definecolor{orange}{rgb}{1,0.5,0}
\definecolor{darkred}{rgb}{.7,0,0}
\definecolor{darkblue}{rgb}{0,0,.6}
\definecolor{grey}{rgb}{.6,.6,.6}
\definecolor{dimgreen}{rgb}{0.2,0.7,0.2}

\newcommand{\jvdomit}[1]{}

\newcommand{\bracketprime}{(\prime)}

\newcommand{\nr}[1]{{\color{darkgreen}{#1}}} 

\usepackage{multirow}
\usepackage{physics}
\usepackage{hyperref}
\hypersetup{colorlinks=true,breaklinks,linkcolor=blue,urlcolor=blue,citecolor=blue}
\usepackage{orcidlink}

\begin{document}

\preprint{}

\title{Compressing local vertex functions from the multipoint numerical renormalization group using quantics tensor cross interpolation}
\author{Markus Frankenbach\,\orcidlink{0009-0002-8907-0031}}
\affiliation{\LMUMunich}

\author{Marc K.\ Ritter\,\orcidlink{0000-0002-2960-5471}}
\affiliation{\LMUMunich}

\author{Mathias Pelz,\orcidlink{0009-0001-3282-6742}}
\affiliation{\LMUMunich}

\author{Nepomuk Ritz\,\orcidlink{0009-0006-5173-5201}}
\affiliation{\LMUMunich}

\author{Jan von Delft\,\orcidlink{0000-0002-8655-0999}}
\affiliation{\LMUMunich}

\author{Anxiang Ge \orcidlink{0009-0002-6603-4310}}
\affiliation{\LMUMunich}

\date{\today}

\begin{abstract}
\medskip
\noindent
    The multipoint numerical renormalization group (mpNRG) is a powerful impurity solver that provides accurate spectral data useful for computing local, dynamic correlation functions in imaginary or real frequencies non-perturbatively across a wide range of  interactions and temperatures. It gives access to a local, non-perturbative four-point vertex in imaginary
    and real frequencies, which can be used as input for subsequent computations such as diagrammatic extensions of dynamical mean--field theory.
    However, computing and manipulating the real-frequency four-point vertex 
    on large, dense grids quickly becomes numerically challenging when the density and/or the extent of the frequency grid 
    is increased. 
    In this paper, we compute four-point vertices in a strongly compressed quantics tensor train format using quantics tensor cross interpolation, starting from discrete partial spectral functions obtained from mpNRG.
    This enables evaluations of the vertex on frequency grids with resolutions far beyond the reach of previous implementations. We benchmark this approach on the four-point vertex of the
    single-impurity Anderson model across a wide range of physical parameters, both in its full form and its asymptotic decomposition.
    For imaginary frequencies, the full vertex can be represented
    to an accuracy on the order of $2\cdot 10^{-3}$ with maximum bond dimensions not exceeding 120. The more complex full real-frequency vertex requires maximum bond dimensions not exceeding 170 for an accuracy of $\lesssim 2\%$.
    Our work marks another step toward tensor-train-based diagrammatic calculations for correlated electronic lattice models starting from a local, non-perturbative mpNRG vertex.
\end{abstract}

\maketitle

\section{Introduction}\label{sec:introduction}
In the study of strongly correlated systems, correlations at the two-particle level play a key role.
A powerful  framework for computing two-particle correlation functions is given by quantum field theory approaches such as the functional renormalization group (fRG) \cite{Metzner2012}
or (closely related \cite{Kugler2017,Kugler2017b,Kugler2018}) the parquet equations \cite{Bickers2004}. While these methods formally provide exact and unbiased equations at the four-point level, solving them in practice requires some approximations. A common choice is the perturbative parquet approximation, which limits the applicability of these methods to weak interactions. In order to apply these diagrammatic methods to correlated electronic lattice systems in the physically relevant strong interaction regime, it has been proposed to combine them with dynamical mean--field theory (DMFT) \cite{Rohringer2018}. DMFT approximates the self-energy to be local, i.e., momentum-independent, thereby neglecting spatial correlations but capturing local correlations non-perturbatively \cite{Georges1996}. In the form of DMF$^2$RG \cite{Taranto2014} or the dynamical vertex approximation (D$\Gamma$A) \cite{Toschi2007, Held2008}, the fRG or the parquet equations can, in principle, be used to self-consistently add non-local correlations on the two-particle level on top of the local DMFT result.

However, such calculations entail two numerical challenges: the solution of the impurity model arising in the self-consistent DMFT loop and, subsequently, solving the fRG or parquet equations for frequency- \textit{and} momentum-dependent vertices. The present work is concerned with the \textit{interface} between these two steps, i.e., the conversion of local four-point spectral functions obtained from an impurity solver to a four-point (4p) vertex.
An impurity solver that yields such 4p spectral functions is the multipoint numerical renormalization group (\mbox{mpNRG})\cite{KuglerMultipointCorrelators,MultipointNRG}.
This extension of the numerical renormalization group (NRG) \cite{Wilson1975, Bulla2008} is capable of computing both imaginary and real-frequency local correlation functions up to the four-point level in the form required for a subsequent diagrammatic extension of DMFT \cite{sIE2024, Ritz2025}. Just as NRG, which has been the gold standard for solving impurity problems on the two-point level for decades \cite{Krishna-murthy1980, Bulla1998}, mpNRG can be applied to a wide range of parameters, including large interactions and low temperatures.
A central ingredient to mpNRG are spectral representations of time-ordered correlation functions in the frequency domain
\cite{KuglerMultipointCorrelators}. These represent correlators as convolutions of formalism-dependent but system-independent kernels with formalism-independent but system-dependent partial spectral functions (PSFs). While the former are known analytically, the latter are obtained from their respective Lehmann representations, using the eigenenergies and (discarded) eigenstates obtained from mpNRG.
The local 4p vertex can be computed
using the symmetric improved estimator (sIE) technique \cite{sIE2024}, which avoids the numerically unstable
amputation of two-point (2p) Green's functions.

An appealing feature of spectral representations is that the same set of PSFs can be used to obtain imaginary- and real-frequency correlation functions.
However, even when energy conservation is exploited, the 4p vertex is a huge, three-dimensional object. Hence its computation from PSFs on a large, dense grid quickly becomes challenging or even unfeasible due to its huge memory footprint.
Furthermore, performing calculations with such vertices as required in fRG or parquet calculations poses a major challenge \cite{Tam2013, AnxiangNepomukKeldysh, NepomukAnxiangKeldyshCode}.

It is thus highly desirable to represent 4p vertices in a \textit{compressed format} that reduces the computational cost of the operations occurring in diagrammatic calculations.
A promising candidate for compression is the quantics tensor train (QTT) representation \cite{Oseledets2009,Khoromskij2011} of multivariate functions, which has recently proven
useful in various areas of physics \cite{Ye2022,Shinaoka2023,Jolly2023,Sroda2024,rohshap2024,Hoelscher2025}. Its first application in the context of many-body theory was in Ref.\ \onlinecite{Shinaoka2023}, a study which demonstrated the compressibility of correlation functions and used QTT-based algorithms to solve the Schwinger-Dyson and Bethe--Salpeter equations.
Furthermore, the QTT representation has been employed successfully in imaginary-frequency parquet calculations for the Hubbard atom and the single\nr{-}impurity Anderson model (SIAM), using the parquet
approximation \cite{rohshap2024}.

These recent developments and the need for efficient representations of 4p vertices motivate this work:
We use mpNRG to compute the local vertex of the SIAM as a function of real and imaginary frequencies in QTT format and investigate its compressibility across a broad range of physical parameters. The reason for studying the SIAM is its natural appearance in a DMFT treatment of the Hubbard model and the fact that it can be solved accurately using (mp)NRG.
To compute the vertex of the SIAM, we employ the quantics tensor cross interpolation (QTCI) algorithm \cite{Oseledets2011,Savostyanov2011,Oseledets2010,Ritter2024}, which iteratively constructs a QTT by sparse sampling of the target function.
This sampling-based interpolation enables evaluation of the mpNRG vertex on grids much larger and much denser than those accessible with the previous state-of-the-art \cite{sIE2024}.
For appropriate error tolerances, the maximum bond dimensions (ranks) of the resulting QTTs are within a range where diagrammatic calculations, such as those presented in Ref.\ \onlinecite{rohshap2024}, should be
feasible, even for real frequencies.

This paper is organized as follows: In \Sec{sec:methods}, we recapitulate how
the 4p vertex of the SIAM can be obtained in imaginary or real frequencies from PSFs.
Additionally, we briefly explain key features of the QTCI algorithm and how it is employed in this work. 
In \Sec{sec:results}, we show that imaginary- and, in particular, real-frequency vertices are representable
by low-rank QTTs within a reasonable error margin.
In the final \Sec{sec:SummaryOutlook}, we provide an outlook on how the results of this work may be used to perform diagrammatic calculations for lattice models
in QTT format.
\vfill\null
\section{Methods}
\label{sec:methods}
This section explains how to compute 4p vertices in QTT format. This is achieved in two steps: First,
we convolve PSFs with formalism-dependent frequency kernels
to obtain correlation functions (cf.\ Ref.\ \onlinecite{KuglerMultipointCorrelators}). In a second step, the sIE scheme is employed \cite{sIE2024} to extract the 4p vertex
from various correlators and self-energies in a numerically stable fashion. The vertex is 
computed both in its asymptotic 
decomposition \cite{KclassesToschi}, which the sIE naturally yields, and in
its `full' form.
\subsection{Partial spectral functions}
\label{subsec:PSFtoCorr}
The input to our calculations is given by the PSFs
\begin{align}
    \label{eq:PSF}
    \mathcal{S}[\opstring](\bomega)=\int \frac{\md^\ell t}{(2\pi)^\ell} \, \me^{\mi \bomega\cdot\boldsymbol{t}} \Big\langle\prod_{i=1}^{\ell}\mathcal{O}^i(t_i)\Big\rangle,
\end{align}
depending on a tuple $\opstring=(\mathcal{O}^1,\ldots,\mathcal{O}^\ell)$ of operators in the Heisenberg picture and $\ell$ frequency arguments
$\bomega=(\omega_1,\ldots,\omega_\ell)$. By $\langle\mathcal{O}\rangle=\Tr[\me^{-\beta H}\mathcal{O}]/Z$, we denote the thermal expectation value, with the partition function $Z = \mathrm{Tr}[e^{-\beta H}]$ at inverse temperature $\beta = 1/T$.
Time translation invariance implies:
\begin{align}
    \label{eq:PSFdirac}
    \mathcal{S}[\opstring](\bomega)=\delta(\omega_{1\cdots\ell}) \, S[\opstring](\bomega),
\end{align}
with the shorthand $\omega_{1\cdots\ell}=\sum_{i=1}^\ell \omega_i$, thus making $S[\opstring]$ a function
of $\ell-1$ independent frequencies.
In this work, we are primarily interested in the case $\ell=4$, i.e., three-dimensional PSFs.

The PSFs carry the formalism-independent information that is specific to the model itself. The \emph{same} set of PSFs can thus be used to compute Matsubara and Keldysh correlators by convolution with formalism-specific kernels.
In this work, PSFs were computed using \mbox{mpNRG} as described in Ref.\ \onlinecite{MultipointNRG}.
This yields PSFs on a discrete, $(\ell-1)$-dimensional logarithmic energy grid for all relevant operator tuples $\opstring$. This yields the representation
\begin{align}
    \label{eq:PSFdelta}
    \specfun(\bomega) = \sum_{\bepsilon} \specfun(\bepsilon) \, \delta(\bomega-\bepsilon),
\end{align}
where the peak weights $\specfun(\bepsilon)$ and energies $\bepsilon$ are obtained as output of the mpNRG computation. The energies $\bepsilon$ are binned into a Cartesian product of logarithmic grids.

\subsection{Matsubara formalism}
A Matsubara correlator $\fullcorr$ depending on $\ell$ operators $(\mathcal{O}^1,\ldots,\mathcal{O}^{\ell})$ can be expressed via $\ell!$ PSFs $S[\opstring_p]$,
as was shown in Sec.\,II.C of Ref.\,\onlinecite{KuglerMultipointCorrelators}:
\begin{align}
    \label{eq:MatsubaraPSFtoG}
    \fullcorr=\sum_{p}\!\! G_p(\mi\bomega_p)=\sum_{p,\bepsilon} \boldsymbol{\zeta}^{p} K(\mi\bomega_p - \bepsilon_p)S[\boldsymbol{\mathcal{O}}_p](\bepsilon_p).
\end{align}
The frequencies $\bomega=(\omega_1,\ldots,\omega_\ell)$ are restricted to discrete fermionic or bosonic
grids, depending on the type of the respective operator $\mathcal{O}^i$.
Similarly to \Eq{eq:PSFdirac}, frequency conservation $\omega_{1\ldots\ell}=0$ is understood. 
The sum $\sum_p$ is over all permutations of $\ell$ elements, permuting frequency arguments
and operators accordingly. Using the shorthand $\overline{i}=p(i)$, we can then write
\begin{alignat}{2}
    \bomega_p
    &=
   \left(\omega_{p(1)},\ldots,\omega_{p(\ell)}\right)
    &&=
    \left(\omega_{\overline{1}},\ldots,\omega_{\overline{\ell}}\right),
    \\
    \boldsymbol{\mathcal{O}}_p
    &=
    \left(\mathcal{O}_{p(1)},\ldots,\mathcal{O}_{p(\ell)}\right)
    &&=  \left(\mathcal{O}_{\overline{1}},\ldots,\mathcal{O}_{\overline{\ell}}\right).
\end{alignat}
Depending on whether $p$ transposes an even or odd number of fermionic operators,
a sign factor $\boldsymbol{\zeta}^p=\pm 1$ is required in \Eq{eq:MatsubaraPSFtoG}.
The summands $G_p(\mi\bomega_p)$ are termed \textit{partial correlators}.
Most importantly, \Eq{eq:MatsubaraPSFtoG} also introduces the Matsubara frequency kernel $K$, which reads
\begin{align}
    \label{eq:MatsubaraKernel}
    K(\boldsymbol{\Omega}_p) =\!
    \begin{cases}
        &\hspace{-0.30cm}
        \prod_{i=1}^{\ell-1} \Omega_{\overline{1} \cdots \overline{i}}^{-1}
        \hfill
        ~\text{if}~
        \prod_{i=1}^{\ell-1}\Omega_{\overline{1} \cdots \overline{i}}\neq 0,
        \\[0.3cm]
        &\hspace{-0.30cm}
        \!-\frac{1}{2}\!\biggl[
            \beta + \sum\limits_{\substack{i=1\\ i\neq j}}^{\ell-1}\Omega_{\overline{1} \cdots \overline{i}}^{-1}
        \biggr]\!
        \prod\limits_{\substack{i=1\\ i\neq j}}^{\ell-1}
        \Omega_{\overline{1} \cdots \overline{i}}^{-1}
        ~\text{if}~
        \exists j:
        \Omega_{\overline{1} \cdots \overline{j}}=0,
    \end{cases}
\end{align}
where $\Omega_j=\mi\omega_j-\epsilon_j$ (cf.\ \Eq{eq:MatsubaraPSFtoG}) and $\Omega_{\overline{1}\ldots\overline{i}}=\mi\omint{i}-\epsint{i}$ with $\mi\omint{i}=\sum_{j=1}^i\mi\omega_{\bar{j}}$ and $\epsint{i}=\sum_{j=1}^i\epsilon_{\bar{j}}$.
The definition \eqref{eq:MatsubaraKernel} assumes that at most one of the partial sums $\Omega_{\overline{1}\ldots\overline{j}}$ vanishes, which is the case if there is at most one bosonic Matsubara frequency (this is always true in the present work). The first case in \Eq{eq:MatsubaraKernel} is called regular kernel, the second case anomalous kernel.
In the most important situation, $\ell=4$ and $\prod_{i=1}^{\ell-1}\Omega_{\bar{1}\ldots\bar{i}}\neq 0$,
the spectral representation of a partial correlator for permutation $p$ is simply given by
\begin{align}
    \label{eq:MatsubaraGpregular}
    G_p(\mi\bomega_p) = 
    \hspace{-1ex} \sum_{\bepsintthree} \hspace{-1ex}
    S[\opstring_p](\bepsintthree)
    \prod_{i=1}^3 (\mi\omint{i}-\epsint{i})^{-1} \nr{.}
\end{align}
\Eq{eq:MatsubaraGpregular} assumes that the PSFs $S[\opstring_p]$ are parametrized in partially summed energies $(\bepsintthree)$, which
is always the case for our data (see Ref.\ \onlinecite{MultipointNRG}).

\subsection{Keldysh formalism}
\label{subsec:keldysh_formalism}
We now turn to the relation between PSFs and correlators in the Keldysh formalism \cite{KeldyshOriginal, SchwingerKeldysh, Kadanoff1962}. For details and derivations, see Sec.\,II.D of Ref.\,\onlinecite{KuglerMultipointCorrelators}.
Keldysh $\ell$-point correlators $G^{\boldsymbol{k}}(\bomega)$ carry a Keldysh index $\boldsymbol{k}=(k_1,k_2,\ldots k_\ell)$ with
$k_i\in\{1,2\}$. Their spectral representation is analogous to \Eq{eq:MatsubaraPSFtoG}:
\begin{subequations}
\begin{align}
    \label{eq:KeldyshPSFtoG}
    \fullcorrK&=\frac{2}{2^{\ell/2}}\sum_{p}G_p^{\boldsymbol{k}}(\bomega_p),\\
    \label{eq:KeldyshGtoGp}
    G_p^{\boldsymbol{k}}(\bomega_p)&=\sum_{\bepsilon}\boldsymbol{\zeta}^p K_b^{\boldsymbol{k}_p}(\bomega_p,\bepsilon_p)S[\opstring_p](\bepsilon_p).
\end{align}
\end{subequations}
It involves $\ell$ real frequencies $\bomega$ that satisfy $\omega_{1\ldots\ell}=0$.
The broadened Keldysh frequency kernel $K_b^{\boldsymbol{k}_p}$ is a linear combination of the broadened, fully retarded kernels $\bretkernel{\lambda}$:
\begin{subequations}
\begin{align}
    \label{eq:keldyshkernel}
    K_b^{\boldsymbol{k}_p}(\bomega_p,\bepsilon_p) &=
\sum_{\substack{\lambda=1\\{k_{\overline{\lambda}}\text{ even}}}}^{\ell}(-1)^{\lambda-1+k_{\overline{1}\ldots\overline{\lambda-1}}}\cdot K_b^{[\lambda]}(\bomega_p,\bepsilon_p),
    \\[0.1cm]
    \bretkernel{\lambda}(\bomega_p,\bepsilon_p) &=
    \prod_{j=1}^{\ell-1}
    \lim_{\gamma_{0} \rightarrow 0^+}
    \int_{\mathbb{R}} \md\omsum{j}' \,
    \frac{\delta_b(\omsum{j}',\epsilon_{\overline{1}\ldots\overline{j}})}{\omsum{j}-\omsum{j}'+\mi\imagshift[j]}
    \, \nr{,} \label{eq:retardedkernel}
    \intertext{{where
$\delta_b(\omega'_{\bar{1}\ldots\bar{j}},\epsilon_{\bar{1}\ldots\bar{j}})
        $ is a broadened version of the Dirac-$\delta$ function appearing in \Eq{eq:PSFdelta}. This broadening ensures a smooth structure of the kernel, free from unphysical poles or $\delta$-peaks. The imaginary shifts \(\mi\imagshift[j]\) in \Eq{eq:retardedkernel} are defined as}
    }
    \label{eq:imagshift}
    \mi\imagshift[j] &=
    \begin{cases}
        \mi\gamma_0\cdot (\ell-j)   & j \geq \lambda,\\
         -\mi\gamma_0\cdot j        & j < \lambda.\\ 
    \end{cases}
\end{align}
\end{subequations}
While the factors $\ell-j$ and $j$ in \Eq{eq:imagshift} can be disregarded in the limit $\gamma_0\rightarrow 0^+$, they remain relevant for the linear broadening. Details on the broadening procedure can be found in \App{app:keldysh_broadening} and Ref.~\onlinecite{MultipointNRG}, Sec.\ VI.

\subsection{Symmetric improved estimators: from correlators to the vertex}
\label{subsec:sIE}
In principle, the one-particle irreducible 4p vertex can be obtained simply by amputating the four external \mbox{2-point} (2p) propagators (``legs'') of the connected impurity Green's function
$G_{\mathrm{con}}[d_{\sigma_1}d^\dag_{\sigma_2}d_{\sigma_3}d^\dag_{\sigma_4}]$ (cf.\ \Sec{subsec:SIAM}). In practice, however, this leads to pronounced numerical artifacts, especially at asymptotically large frequencies, where both functions decay to zero.
A numerically stable scheme that avoids direct amputation is the symmetric improved estimator (sIE) technique introduced in Ref.~\onlinecite{sIE2024}. In addition, this
method yields the vertex in its asymptotic decomposition \cite{KclassesToschi}. This decomposition separates the contributions that decay only in one or two frequency directions from the genuinely three-dimensional core vertex 
$\gamcore$, which asymptotically decays in all three frequencies,
\begin{equation}
\begin{aligned}
    \label{eq:Kclasses}
    \Gamma(\omega,\nu,\nu') &=
        \gamcore(\omega,\nu,\nu')\\
        &\quad + \sum_{r=a,p,t} {\left[
            \kclass_2^r(\omega_r,\nu_r) +
            \kclass_{2'}^r(\omega_r,\nu_r') +
            \kclass_1^r(\omega_r)
        \right]}
        \\&\quad + \gambare.
\end{aligned}
\end{equation}
The functions $\kclass_1^r$, $\kclass_2^r$ and $\kclass_{2'}^r$ (not to be confused with the kernels \(K\) defined above) only depend on one or two frequencies if parametrized
in their native channel $r$. They are one- and two-dimensional contributions to the two-particle reducible vertex in channel $r$. This channel can be the  antiparallel ($a$), parallel ($p$), or the transverse ($t$) channel.
These are also known as the particle-hole, particle-hole crossed and particle-particle channels, respectively \cite{sIE2024}.
By $\gambare$ we denote the frequency-independent bare vertex.
Note that the sIE method does not provide a decomposition of $\gamcore$ into two-particle reducible contributions $\kclass_3^{r}$ and a two-particle irreducible term.

In this work, we consider the single-impurity Anderson model with interaction
\(
    H_{\text{int}}=U d^\dagger_{\uparrow} d_{\uparrow} d^\dagger_{\downarrow} d_{\downarrow}
\), where \(d^\dagger_\sigma\) creates an electron with spin \(\sigma\) on the impurity, see Sec.\,\ref{sec:SIAM} for details.
We denote the self-energy of the impurity Green's function $G[d_\sigma,d_{\sigma'}^\dag](\nu)$ as \(\Sigma^{\sigma\sigma'}(\nu)\).
In the absence of a magnetic field, it satisfies \(\Sigma^{\sigma\sigma'}(\omega) = \delta^{\sigma\sigma'} \Sigma(\omega)\).
Following Ref.~\onlinecite{sIE2024}, \(\gamcore\) can be obtained as
\begin{align}
    \label{eq:gamcore}
    \gamcore(\bomega)
    =\quad
    \sum_{\mathclap{a_i\in \{d, q\}}}
 \quad &Y_{a_1}(\omega_1) Y_{a_3}(\omega_3)
        \,G_{\text{con}}[a_1, a_2^\dagger, a_3, a_4^\dagger](\bomega)\nonumber\\
        &\cdot Y_{a_2}(\omega_2)Y_{a_4}(\omega_4),
\end{align}
where we introduced an auxiliary operator \(q = [d, H_{\text{int}}]\), and
\begin{equation}
    \label{eq:Ysymbol}
        Y_{a_i}(\omega_i) =
    \begin{cases}
        -\Sigma\left((-1)^{i-1}\omega_i\right)
        & a_i = d,
        \\
        X=
        \left(
        \begin{smallmatrix}
            0 & 1\\
            1 & 0
        \end{smallmatrix}
        \right)
        & a_i = q,
    \end{cases}
\end{equation}
is a $2\!\times\!2$ matrix acting on the $i$th Keldysh index of $G_{\mathrm{con}}$.
In the Matsubara formalism, $X$ is replaced by scalar unity.
In practice, the main workload in computing $\gamcore$ at a given frequency $\bomega$ is the evaluation of all $2^4=16$ connected correlators, each comprised of $4!=24$ partial correlators.
The quantities $\kclass_1^r$ and $\kclass_{2^{\bracketprime}}^r$ can be computed using an analogous approach presented in \App{app:K1K2}.
\subsection{Quantics Tensor Cross Interpolation} \label{subsec:QTCI}
Having summarized the evaluation of the $4$-point vertex
in its asymptotic decomposition, we next discuss the quantics tensor cross interpolation (QTCI) method \cite{Oseledets2011,Oseledets2010,Savostyanov2011,Ritter2024}, which we used 
to obtain vertex functions in the form of QTTs. Recently, it has been shown in the Matsubara formalism that this representation is well-suited for efficient diagrammatic calculations \cite{rohshap2024}. 
We discuss only the basics of QTCI here. For a detailed introduction, we refer to Ref.\ \onlinecite{Fernandez2024}.

Let us begin with the quantics representation \cite{Oseledets2009,Khoromskij2011}. Consider a one-dimensional function $f(\omega)$ defined on a discrete, equidistant
grid $\{\omega_0,\ldots,\omega_{2^R-1}\}$ consisting of $2^R$ points with $\omega_m\in\mathbb{R}$.
The grid index $m$ of a point $\omega_m$ can be written in binary representation
\begin{align}
    m = \sum_{\ell=1}^{R}2^{R-\ell}\qbit,\; \qbit\in\{0,1\},
\end{align}
so that $m$ can be identified with the $R$-tuple $(\sigma_1,\ldots,\sigma_R)$.
Hence\nr{,} the mapping $m\mapsto f(\omega_{m})$ can be viewed as an $R$-leg tensor
$F_{\sigma_1\cdots\sigma_R}=f(\omega_{m(\{\sigma_\ell\})})$.
This so-called quantics encoding can be generalized to higher-dimensional functions, in particular to
functions $f(\omega,\nu,\nu')$ of three frequency arguments.
The frequencies lie on a Cartesian product of 1D grids, each of size $2^R$.
We use a binary encoding
\begin{align}
    (\omega_i,\nu_j,\nu'_k)=\big((\sigma_{11},\ndots,\sigma_{1R}), (\sigma_{21},\ndots,\sigma_{2R}), (\sigma_{31},\ndots,\sigma_{3R})\big),
\end{align}
with the binary variables $\sigma_{n\ell}$ labelled by $n=1,2,3$ for $\omega,\nu,\nu'$. The function $f$ can then be represented by a tensor with $3R$ indices:
\begin{align}
    \label{eq:3dquantics}
    F_{\boldsymbol{\sigma}}=F_{\sigma_{11}\sigma_{21}\sigma_{31}\cdots\sigma_{1R}\sigma_{2R}\sigma_{3R}}=f(\omega_i,\nu_j,\nu'_k).
\end{align}
Importantly, note that the tensor indices in \Eq{eq:3dquantics} have been \textit{interleaved}, such that the indices corresponding to the same length scale $2^{R-\ell}$ in different variables \(\sigma_{1\ell}, \sigma_{2\ell}, \sigma_{3\ell}\), are adjacent.
Alternatively, triples ($\sigma_{\ell1},\sigma_{\ell2},\sigma_{\ell3}$) of legs can be fused to
single legs $\tilde{\sigma}_\ell=\sum_{n=1}^{3}2^{n-1}\sigma_{\ell n}$, which yields the \textit{fused} representation of $f$ as an $R$-leg tensor:
\begin{equation}
    \label{eq:3dquantics_fused}
    \widetilde{F}_{\tilde{\boldsymbol{\sigma}}}=\widetilde{F}_{\tilde{\sigma}_1\cdots\tilde{\sigma}_R}=f(\omega_i,\nu_j,\nu'_k).
\end{equation}

The second ingredient of QTCI is the tensor cross interpolation (TCI) algorithm \cite{Oseledets2011,Oseledets2010,Savostyanov2011,Fernandez2024},
which approximates tensors $F_{\bsigma}=F_{\sigma_1\cdots\sigma_L}$ (with $L=R,2R$ or $3R$ for one-, two- or three-dimensional functions, respectively) using tensor trains
constructed from a sampled subset of all tensor elements. If a low-rank factorization of the tensor exists, the 
number of samples taken is much smaller than the number of elements of the full tensor. This way, the cost of generating all tensor elements, exponentially large in \(R\), can be avoided.

More precisely, the TCI algorithm seeks to find a tensor train (TT)
\begin{equation}
    F^{\text{QTCI}}_{\sigma_1 \ldots \sigma_L} =
\sum_{\alpha_1 \ldots \alpha_{L-1}} [M_1^{\sigma_1}]_{1\alpha_1} [M_2^{\sigma_2}]_{\alpha_1 \alpha_2} \cdots [M_L^{\sigma_L}]_{\alpha_{L-1}1}
\end{equation}
that minimizes the elementwise error
\begin{align}
    \label{eq:tciconvergence}
    \error_{\bsigma}[F]
    =
    \frac{|F^{\mathrm{QTCI}}_{\bsigma} - F_{\bsigma}|}{\max_{\bsigma'}|F_{\bsigma'}|}\,.
\end{align}
Here, the $\alpha_\ell$ are virtual bond indices with $\ell$-dependent bond dimensions, $\alpha_\ell=1,\ndots,\chi_\ell$. The maximum bond dimension, $\chi=\max{\chi_\ell}$, is called the \textit{rank} of $F_{\boldsymbol{\sigma}}^{\mathrm{QTCI}}$. The maximum in \Eq{eq:tciconvergence} is estimated using all sampled entries of $F_{\bsigma}$.
The TCI algorithm optimizes the tensors \([M_\ell^{\sigma_\ell}]_{\alpha_{\ell-1}\alpha_\ell}\) iteratively, progressively sampling \(F_{\bsigma}\), until no \(\bsigma\) is found where the error \(\error_{\bsigma}\) exceeds a given tolerance \(\tcitol\).
During this process, the bond dimensions $\chi_\ell$ are increased dynamically to improve the accuracy of the tensor train representation. 
Finding a tensor train representation of $F^{\mathrm{QTCI}}_{\bsigma}$ with TCI has a computational cost of $\order{R\chi^3}$.

In conjunction with the quantics representation, TCI can be employed to approximate not only tensors, but also functions defined on discrete grids by tensor trains. Once the bond dimensions are saturated, i.e., no longer increase with $R$, the computational cost of QTCI scales linearly in $R$. This translates to an exponential resolution of the target function at linear cost. 
Of course, the function is only approximated within an error margin given by \Eq{eq:tciconvergence}.
The next section details how we used QTCI to compress 4p vertex functions on exponentially large grids.

\subsection{Implementation details}
To obtain the core and full vertices in the Matsubara and Keldysh formalisms as QTTs, we
apply QTCI to functions that evaluate $\gamcore(\bomega)$ and $\Gamma(\bomega)$ on individual frequency points \(\bomega\) to be specified on demand by the TCI algorithm. The frequencies $\bomega$ reside on an equidistant grid of $2^{3R}$ points. For Matsubara grids, the grid spacing is set by the temperature, and the extent of the grid can be increased exponentially by increasing $R$. 
For Keldysh vertices, which are functions of continuous frequencies, one may exponentially increase either the density of grid points, or the extent of the grid, or both, by increasing $R$.
It is important not to precompute the vertices on a dense grid, as this precomputation step would incur costs scaling as \(\order{2^{3R}}\). By avoiding precomputation, \(R\) can be increased to yield grid sizes and/or grid densities beyond those attainable by conventional means. The TCI algorithm samples a \textit{sparse} set of \(\order{\chi^2 R}\) points \(\bomega\), which is much smaller than $2^{3R}$,
the total number of grid points for the 4p vertex functions considered in this work. For this application, function evaluation during sampling is the dominant cost as opposed to the \(\order{\chi^3 R}\) cost of computing prrLU factorizations (see \cite[Sec.\ 3.3]{Fernandez2024}).
More specifically, the computational effort is dominated by the evaluation of partial 4p correlators.
These enter $\gamcore$ and $\Gamma$ via the full correlators appearing in \Eq{eq:gamcore}.
In this section, we discuss how our code evaluates partial correlators in an efficient way. Readers only interested in our results on the compressibility of Matsubara and Keldysh vertices can move on to \Sec{sec:results}.

\subsubsection{Matsubara vertices}

In the Matsubara case, evaluating the regular part of the 4p correlator, \Eq{eq:MatsubaraGpregular}, constitutes the majority of computational cost.
Using the shorthand $\kernelmat=\mi\Omint{i}=(\mi\omint{i}-\epsint{i})^{-1}$, we can rewrite \Eq{eq:MatsubaraGpregular} as 
\begin{align}
    \label{eq:MatsubaraGpregular2}
    G_p(\mi\bomega_p)
    =\!
    \hspace{-1ex}
    \sum_{\bepsintthree}
    \hspace{-1ex}
    k_{\omega_{\bar{1}}\epsilon_{\bar{1}}}
    k_{\omega_{\bar{1}\bar{2}}\epsilon_{\bar{1}\bar{2}}}
    k_{\omega_{\bar{1}\bar{2}\bar{3}}\epsilon_{\bar{1}\bar{2}\bar{3}}}
    \pspecfun(\bepsintthree).
\end{align}
Since $(\bomintthree)$ and $(\bepsintthree)$ live on finite frequency grids, the kernels $\kernelmat$ can be viewed as matrices.
A typical grid size for the spectral function peaks $(\bepsintthree)$ is $70\times70\times70$, while the largest grids for Matsubara frequencies \((\bomintthree)\)
used in this work have $2^{12}=4096$ points in each dimension.
Thus $\kernelmat$ can be precomputed and stored for all relevant grid sizes.

We implemented two methods to speed up the threefold contractions in \Eq{eq:MatsubaraGpregular2}: (i) compressing the kernels and (ii) performing one kernel
contraction as a preprocessing step.

\begin{figure}
    \centering
\includegraphics[width=\columnwidth, trim=0.45cm 0.4cm 0.3cm 0, clip]{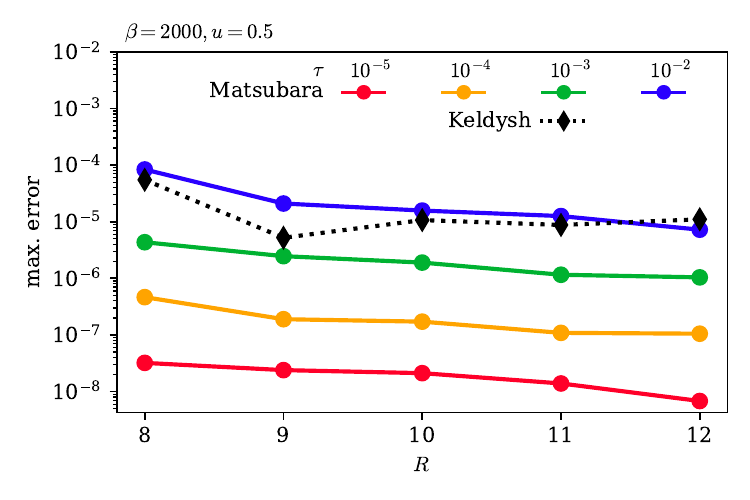}
    \caption{Maximum error of Matsubara (circles) and Keldysh (diamonds) core vertex evaluations with SVD truncations as described in the main text. The error is measured relative to the maximum of the respective vertex function. By $\tau$ we denote the target TCI tolerance, and use a cutoff of $S_{\text{cut}}=10^{-2}\tau$ for Matsubara. For Keldysh, we choose an SVD cutoff of $10^{-6}$ times the largest singular value, and find that this yields results that are sufficiently accurate for a tolerance of $\tcitol=10^{-3}$. 
    We show maximum errors over 64000 sampling points for the Matsubara core vertex and $2\cdot10^6$ sampling points for the more complicated Keldysh core vertex. All errors are well below the respective target tolerance $\tau$.} 
    \label{fig:accuracy}
\end{figure}

(i) Compressing the kernels is the more general of the two methods, in that it has a smaller memory footprint ($<1$\,GB per full correlator $G(\mi\bomega)$ for $R=12,\tcitol=10^{-3}$).
The idea is to compress the kernels $k_{\omega\epsilon}$ by exploiting their low-rank structure \cite{SVDdecayMaths,SVDdecayShinaoka}:
We SVD-decompose each $k_{\omega\epsilon}$ and discard singular values below a given cutoff $\sigmacut$, resulting in the approximation
\begin{align}
    \label{eq:kernelsvdMatsubara}
    k_{\omega\epsilon} \approx \sum_a U_{\omega a} S_{a} V^{\dag}_{a \epsilon}.
\end{align}
We then contract the singular values $S_{a}\geq\sigmacut$ and the right-hand isometries $V^\dag$ with the PSF by performing $\epsilon$ sums to obtain
a smaller rank-3 tensor $A$, thus reducing the cost of the threefold summation in \Eq{eq:MatsubaraGpregular2}:
\begin{subequations}
\begin{align}
    \label{eq:MatsubaraGpImplementation}
    G_p(\mi\bomega_p)
    &=
    \sum_{\mathclap{a_1 a_2 a_3}}
    A_{a_1 a_2 a_3} \prod_{i=1}^3  \isokernel,\\
    \label{eq:MatsubaraAtensor}
    A_{a_1a_2a_3}
    &=
    \sum_{\mathclap{\bepsintthree}}
    S[\opstring_p](\bepsintthree)
    \prod_{i=1}^3
    S_{a_i} V^{\dag}_{a_i\epsilon_{\overline{1}\cdots\overline{i}}}
    \,.
\end{align}
\end{subequations}
The computations \eqref{eq:kernelsvdMatsubara} and \eqref{eq:MatsubaraAtensor} are performed during preprocessing prior to the QTCI run.
Note that this treatment of the low-rank Matsubara kernels is closely related to the so-called intermediate representation (IR) of Matsubara Green's functions, see Refs.~\cite{Dirnboeck2024,IRreference}.
The cutoff $\sigmacut$ should be chosen as to introduce an error significantly below the
TCI tolerance $\tcitol$ in all target quantities.
While one can bound the error in \Eq{eq:MatsubaraGpImplementation}, e.g., using
the Cauchy--Schwarz inequality, these estimates were found to be very conservative. 
We observed that setting $\sigmacut=10^{-2} \, \tcitol$ leads to errors more than two orders of magnitude
below the TCI tolerance when evaluating correlators and vertices. This is shown in \Fig{fig:accuracy}, where we plot the accuracy of Matsubara (and Keldysh) core vertex evaluations for different TCI tolerances $\tau$ and numbers of quantics bits $R$. In Matsubara, the accuracy improves with increasing $R$. This is because, for a fixed SVD cutoff, fewer singular values are discarded for larger $R$.

A further speedup can be achieved by realizing that, even though all singular values $S_{a}$ in \Eq{eq:kernelsvdMatsubara}
are larger than $\sigmacut$, their products appearing in \Eq{eq:MatsubaraAtensor} can become
negligibly small. Ordering $S_{a_i}$ by decreasing magnitude, we therefore discard all entries
$A_{a_1a_2a_3}$ where $a_1+a_2+a_3$ is larger than some integer $N$:
\begin{align}
    G_p(\mi\bomega_p)
    =
    \sum_{\mathclap{\sum_i a_i\leq N}}
    A_{a_1 a_2 a_3}
    \prod_{i=1}^3\isokernel
    \,.
\end{align}
A similar truncation is useful when constructing IR 4pt Green's functions, see Ref.\ \onlinecite{WallerbergerBSE}. To determine $N$ for a prescribed tolerance $\tau$, we estimate
the contribution from terms with $\sum_ia_i>N$ via the Cauchy--Schwarz inequality:
\begin{align}
    \label{eq:HoelderEstimate}
    & \left|
        \sum_{\sum_i a_i>N} \hspace{-0.85em} A_{a_1a_2a_3} \prod_{i=1}^3\isokernel 
    \right|
    \nonumber\\
    &\qquad \leq
    \sqrt{\sum_{\sum_ia_i>N} \hspace{-0.85em} |A_{a_1a_2a_3}|^2}\,
    \prod_{i=1}^3 \max_{\omega_{i}}
    \sqrt{\sum_{a_i}\left|U_{\omega_ia_i}\right|^2} \nonumber\\
    &\qquad =\sqrt{\sum_{\sum_ia_i>N} \hspace{-0.85em} |A_{a_1a_2a_3}|^2}\;.
\end{align}
The second factor in the second line of \Eq{eq:HoelderEstimate} is equal to one,
since the $U$'s are isometries. Hence, \Eq{eq:HoelderEstimate} provides a simple bound on the error
in $G_p(\mi\bomega)$, which is independent of the frequency at hand. We choose $N$ such that
\begin{align}
\label{eq:tucker_cut_criterion}
\sqrt{ \sum_{\sum_ia_i>N}|A_{a_1a_2a_3}|^2 }
\leq
\frac{\tcitol}{10} \max_{\bomega}|G(\mi\bomega)|,
\end{align}
which ensures an error one order of magnitude below the TCI tolerance in the full correlator $G$.
While this error in principle occurs \textit{per} partial correlator $G_p$, the criterion \eqref{eq:tucker_cut_criterion} was observed to yield sufficient accuracy. Overall, this first method gives a substantial speedup compared to directly performing the contractions in Eq.\,\ref{eq:MatsubaraGpregular2}:
It accelerates the evaluation of the full 4p impurity correlator $G[d_{\uparrow},d_{\uparrow}^\dag,d_{\uparrow},d_{\uparrow}^\dag]$ at $\beta=2000,\,u=0.5$ (cf.\ \Sec{subsec:SIAM})
at a single frequency in an $R=12$ quantics grid using an SVD cutoff of $\sigmacut = 10^{-5}$ by more than a factor 60.
This observation simply reflects the strong compresssibilty of the Matsubara kernels.

(ii) A more straightforward way to speed up pointwise evaluations of partial correlators \eqref{eq:MatsubaraGpregular2} is to precompute
one of the three contractions before running QTCI. This yields an object depending on the variables $(\omega_{\bar{1}},\epsinttwo,\epsintthree)$,
\begin{equation}
    \label{eq:Bpintermediate}
    B_p(\omega_{\bar{1}},\epsinttwo,\epsintthree)=\sum_{\epsilon_{\bar{1}}}k_{\omega_{\bar{1}}\epsilon_{\bar{1}}}\pspecfun(\bepsintthree).
\end{equation}
$B_p$ gives access to $G_p$ via
\begin{equation}
    G_p(\mi\bomega_p)
    =
    \sum_{\epsinttwo\epsintthree}
    k_{\omega_{\bar{1}\bar{2}}\epsinttwo}
    k_{\omega_{\bar{1}\bar{2}\bar{3}}\epsintthree}
    B_p(\omega_{\bar{1}},\epsinttwo,\epsintthree).
\end{equation}
In this approach, we only have two kernel contractions in each evaluation $G_p$, but have to store the intemediates \eqref{eq:Bpintermediate} for all partial correlators.
Their size grows linearly in the grid frequency grid size, i.e., as $2^R$ where $R$ is the number of quantics bits.
For example, if $R=12$ and the PSFs live on a $\threetimes{70}$ logarithmic grid (which results from $2\times6$ decades of energy bins with 8 points per decade and discarding zeros in the PSFs), each full correlator consumes $12.9$\,GB of memory.
On an $R=12$ grid at $\beta=2000,\,u=0.5$, the precomputation also yields a speedup of about a factor 60. But in contrast
to the compression of kernels (\Eq{eq:kernelsvdMatsubara}), this speedup is independent of the TCI tolerance \(\tcitol\).
Overall, method (ii) is recommended as long as its memory demands can be met, because it evaluates correlators in a numerically exact fashion.

\subsubsection{Keldyh vertices}
\label{subsec:keldysh_implementation_details}
In the Keldysh formalism, evaluating partial correlators $G_p^{\boldsymbol{k}}$ (cf.\ \Eq{eq:KeldyshGtoGp}) also comes down to threefold contractions
of a \mbox{3-dimensional} PSF with kernel matrices. This can be seen by rewriting the kernel $\bretkernel{\lambda}$ \eqref{eq:retardedkernel} as
a product of one-dimensional kernels evaluated at frequencies $\omint{i}$:
\begin{subequations}
    \begin{align}
    \label{eq:keldysh1Dkerneldef}
    \bretkernel{\lambda}(\bomega_p,\bepsilon_p)&=\prod_{i=1}^{3}k^{[\lambda,i]}_b(\omint{i},\epsint{i}),\\
    k^{[\lambda,i]}_b(\omint{i},\epsint{i})&=\lim_{\gamma_{0} \rightarrow 0^+}
    \int_{\mathbb{R}} \md\omsum{i}' \,
    \frac{\delta_b(\omsum{i}',\epsilon_{\overline{1}\ldots\overline{i}})}{\omsum{i}-\omsum{i}'+\mi\imagshift[i]}.
    \end{align}
\end{subequations}
Equation (\ref{eq:KeldyshGtoGp}) can then be written as:
\begin{subequations}
\begin{align}
    G_p^{\boldsymbol{k}}(\bomega_p)=&\sum_{\substack{\lambda=1\\ k_{\overline{\lambda}} \text{even}}}^4 (-1)^{\lambda-1+\kelsum{\lambda-1}}
    \cdot \retgp(\bomega_p),\\\nonumber
    \label{eq:KeldyshContraction}
    \retgp(\bomega_p)&=\sum_{\bepsintthree}\pspecfun(\bepsintthree)\times\\
&\qquad\qquad\quad\;
\prod_{i=1}^{3}k^{[\lambda,i]}_b(\omint{i},\epsint{i}).
\end{align}
\end{subequations}
The ensuing contractions \eqref{eq:KeldyshContraction}
to be performed for $\lambda=1,\ndots,4$ are analogous to \Eq{eq:MatsubaraGpregular2}.
However, interpolating the complex structure of the Keldysh vertex requires more evaluations compared to its Matsubara counterpart.
At the same time, the memory cost of a precomputation analogous to \Eq{eq:Bpintermediate} becomes prohibitive for large ($R\gtrsim 12$) frequency grids, since it
must be applied to $\retgp$ for $\lambda=1,\ndots,4$ and for each partial correlator.
For these reasons, the optimization of \Eq{eq:KeldyshContraction} needs to go beyond the compression scheme for the Matsubara kernels from
Eqs. \eqref{eq:kernelsvdMatsubara} and \eqref{eq:MatsubaraAtensor}.
To this end, we exploit the fact that the structure of the 1D kernels $k^{[\lambda,i]}_b(\omint{i},\epsint{i})$ becomes
simpler at large frequencies $\omint{i}$:
We divide the $\omint{i}$ grid into $n_L$ equally-sized intervals $I_1^i,\ldots,I_{n_L}^i$, with $n_L=2^3$ as a default. 
Then, for each dimension $i$ and each interval $I^i_j$, we SVD-decompose the restricted kernel
\begin{align}
    k^{[\lambda,i]}_b(\omint{i},\epsint{i})\bigg|_{\omint{i}\in I^i_j}\approx \sum_{a_i} U_{\omint{i}a_i}^{ij}S^{ij}_{a_i} V^{\dag ij}_{a_i\epsint{i}},
\end{align}
discarding singular values that are at least 6 orders of magnitude smaller than the largest singular value.
This strategy of partitioning the $\omint{i}$ grid prior to the SVD truncation allows us to discard more singular values in outer intervals, where the kernel is more compressible.
Next, for each triple of intervals $(I^1_{k},I^2_{l},I^3_{m})$, we contract the corresponding singular values
and right hand isometries into the PSF. While this entails precomputing $n_L^3$ 3-leg tensors of the form
\begin{align}
    &\left(A^{klm}\right)_{a_1a_2a_3}
    =\nonumber\\=&
    \sum_{\mathclap{\bepsintthree}}(SV^\dag)_{a_1\epsilon_{\bar{1}}}^{1k}(SV^\dag)_{a_2\epsinttwo}^{2l}(SV^\dag)_{a_3\epsintthree}^{3m}
    S[\opstring_p](\bepsintthree),
\end{align}
{it yields a substantial speedup in evaluations of $G^{\boldsymbol{k}}_{}(\bomega)$: For $\beta=2000$, $u=0.5$, $\omega_{\max}=0.65$ (cf. \Sec{subsec:SIAM}) and $R=12$ this scheme 
is about a factor 150 faster than a naive kernel contraction. This speedup refers to an average over $2\cdot10^{5}$ evaluations on random frequency points, since the compressibility of the kernels depends on the intervals $(I^1_k,I^2_l,I^3_m)$ the frequencies $(\omega_{\bar{1}},\omega_{\bar{1}\bar{2}},\omega_{\bar{1}\bar{2}\bar{3}})$ belong to. Indeed, truncated isometries $U^{ij}$ pertaining to the outermost intervals usually have about 5 times fewer rows than those of the inner intervals.
That Keldysh core vertex evaluations using the above scheme are sufficiently accurate (i.e.\ to more than $10^{-3}$, see \Sec{subsec:KeldyshResults}) is verified in \Fig{fig:accuracy}.

Having explained the optimization of vertex evaluations, we turn to the settings chosen in the QTCI routine.
All of our code is written in Julia (versions 1.9.4 and 1.10.3), using the TCI package \texttt{TensorCrossInterpolation.jl}, the quantics utilities \texttt{QuanticsGrids.jl} as well as the QTCI
package \texttt{QuanticsTCI.jl} of the tensor4all collaboration \cite{tensor4all,Fernandez2024}.
The latter exposes the \texttt{quanticscrossinterpolate} routine, which is the entry point of the QTCI algorithm and offers various settings:
We used the default \texttt{:backandforth} sweep strategy and the \texttt{:fullsearch} pivot  search strategy.
The increase in computational cost entailed by a full pivot search was accepted to ensure a reliable interpolation.
For Matsubara objects, the interleaved representation was chosen to obtain maximum memory compression. In the Keldysh case, the interleaved representation
exhibited convergence problems: After 80 sweeps (with $\leq5$ sweeps until convergence being common), a QTT with an error significantly exceeding the tolerance
was obtained. Switching to the fused representation solved this problem. This is due to the fact that a 2-site update in a 3D fused representation corresponds
to a 6-site update in the interleaved representation, which implies more extensive sampling of the target function.
Another choice worth mentioning is that of initial pivots:
For Keldysh vertices, it was sufficient to use the grid center as the only initial pivot.
In Matsubara, the same choice occasionally lead to premature termination of the TCI algorithm, resulting in a QTT representation that was missing relevant features.
We therefore chose 125 initial pivots forming a cube at Matsubara frequencies $(\omega_{i},\nu_{j-1},\nu'_{k-1})$ with
$i,j,k\in\{-2,\ndots,2\}$. On fermionic grids, the cube is thus centered around $\nu_{-1}=\nu'_{-1}=-\pi T$.
This choice of initial pivots ensures that the sharp Matsubara vertex structure around the origin is properly sampled.
Finally, since vertex evaluations are the bottleneck of our QTCI-compressions,
a significant speedup can be achieved via multithreading.
The samples evaluated during a two-site optimization step (see Ref.\ \cite[Sec.\ 4.3]{Fernandez2024}) are independent of one another, and can therefore be evaluated in parallel.

We tested our code for evaluating vertex functions with the sIE scheme against the prior Matlab implementation used in Ref.\ \onlinecite{sIE2024}.
We found numerically exact agreement with a normalized discrepancy \(<10^{-13}\) for the Matsubara quantities. In Keldysh the maximum discrepancy in the core vertex between our Julia code and the Matlab code of Ref.\ \onlinecite{sIE2024} is about $0.002\cdot||\gamcore||_{\infty}$ (with the supremum norm $||\cdot||_{\infty}$). This discrepancy can be attributed to small differences in the broadening implementation, mainly the interpolation of the broadened kernel from a logarithmic to a linear grid (cf.\ \App{app:keldysh_broadening}). This discrepancy is one order of magnitude smaller than the error introduced by the arbitrariness inherent in the choice of broadening parameters.
As an additional test, our code was used to generate Keldysh vertex data to check the fulfillment of exact diagrammatic relations of mpNRG data \cite{Ritz2025}.

To conclude this section, \Fig{fig:kernelSVD} compares the singular value spectra of regular Matsubara kernels (\Eq{eq:MatsubaraKernel}) and broadened, fully retarded Keldysh kernels (\Eq{eq:keldysh1Dkerneldef}) at different temperatures. As expected, the singular values of both Matsubara and Keldysh kernels decay significantly faster at higher temperatures. In Keldysh, this is due to the temperature-dependent linear broadening $\gamma_{\mathrm{L}}$ (see \App{app:keldysh_broadening}). Moreover, the singular values of Keldysh kernels decay much more slowly than their Matsubara counterparts at the same temperature. This reflects the more complex structure, i.e., lower compressibility, of Keldysh vertices.
\begin{figure}[tbp]
\centering
\includegraphics[width=\columnwidth, trim=0.2cm 0.4cm 0.2cm 0, clip]{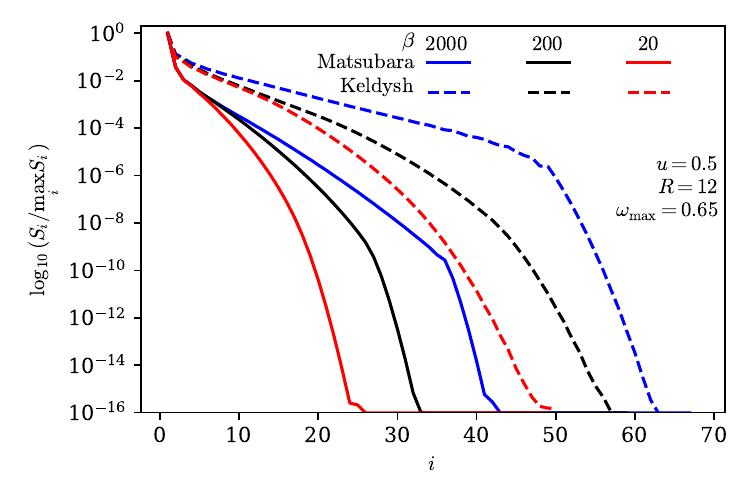}
\caption{
Singular values $S_i$ of regular Matsubara kernels (\Eq{eq:MatsubaraKernel}, solid line) and broadened, fully retarded Keldysh kernels (\Eq{eq:keldysh1Dkerneldef}, dashed line). We show kernels at inverse temperatures $\beta\in\{20,200,2000\}$ and interaction $u=0.5$. The frequency grids are bosonic with $2^{12}$ points, with the Keldysh grid ranging from $-0.65$ to $\omega_{\max}=0.65$.
}
\label{fig:kernelSVD}
\end{figure}

\section{Results}
\label{sec:results}
In this section, we show how QTCI performs in compressing the 4p vertex of the single-impurity Anderson model (SIAM)
in the Matsubara (\Sec{subsec:MatsubaraResults}) and Keldysh (\Sec{subsec:KeldyshResults}) formalisms. We discuss the benefits of its QTT representation compared to storing the vertex on dense frequency grids, considering both the core and the full vertex ($\gamcore$ and $\Gamma$ in \Eq{eq:Kclasses}). The asymptotic contributions
are discussed in \App{app:K1K2}. The two most relevant numerical parameters are the number $R$ of quantics bits,
corresponding to a grid with $2^R$ points in each dimension and the maximum bond dimension $\chi$, which serves as a measure for compressibility. 
\begin{table}[tbp]
    \setlength{\tabcolsep}{5pt}
    \centering
    \begin{tabular}{ccc|cc}
       $U$ & $\Delta$ & $u$ &  $T_K$ & $\beta_K=1/T_K$\\
       \midrule
       0.05 & 0.0318 & 0.5 & $4.14\cdot 10^{-2}$ & 24.2 \\
       0.05 & 0.0159 & 1.0 & $9.58\cdot 10^{-3}$ & 104\\
       0.05 & 0.0106 & 1.5 & $3.57\cdot 10^{-3}$ & 280 \\
       0.05 & 0.00530 & 3.0 & $3.36\cdot 10^{-4}$ & 2980 \\
       0.05 & 0.00318 & 5.0 & $2.06\cdot 10^{-5}$ & 48400\\
    \end{tabular}
    \caption{Kondo temperatures $T_K=T_K(U,\Delta)$ with inverses $\beta_K=1/T_K$ for different parameter sets. The Kondo temperature was computed via the Bethe ansatz solution of the SIAM, see, e.g., Ref.\ \onlinecite{Filippone2018}. All quantities have been rounded to three significant digits.}
    \label{tab:kondo-temperatures}
\end{table}
\subsection{Single impurity Anderson model}\label{sec:SIAM}
\label{subsec:SIAM}
The Hamiltonian of the single impurity Anderson model (SIAM) \cite{SIAM} reads
\begin{equation}
\begin{aligned}
    \label{eq:SIAM}
    H &= \sum_{\sigma} \epsilon_d\,n_\sigma + U n_{\uparrow}n_{\downarrow} + \sum_{k\sigma} \epsilon_k c^\dag_{k\sigma}c_{k\sigma}\\
    &\quad + \sum_{k\sigma} V_k\left(d^\dag_\sigma c_{k\sigma} + \mathrm{h.c.}\right),\quad n_\sigma = d^\dag_\sigma d_\sigma
\end{aligned}
\end{equation}
where $d^\dag_\sigma$ with spin $\sigma\in\{\uparrow, \downarrow\}$ creates an electron in an interacting, single-orbital impurity. $\smash{c^\dag_{b\sigma}}$
creates an electron in a noninteracting bath, coupled to the impurity via a hybridization term $V_{k}$. Electrons on the impurity site interact with the interaction strength $U$.
Since the $c_{k\sigma}$ electrons occur only quadratically, they can formally be integrated out, yielding a frequency-dependent hybridization function $\Delta(\nu)$ as an additional quadratic term for the $d$ electrons.
We choose the hybridization function as
\begin{align}
    \Delta(\nu) = \frac{\Delta}{\pi} \, \ln\left|\frac{\nu + D}{\nu - D}\right| - i\Delta \, \theta(D-|\nu|) ,
\end{align}
with a box-shaped imaginary part,
characterized by the bandwidth $2D$ and the hybridization strength $\Delta\in\mathbb{R}$.
Moreover, we set $\epsilon_d=- U / 2$, which leads to a particle-hole symmetric Hamiltonian.

In the following, energy, temperature and frequencies are measured in units of half the bandwidth $D=1$.
The interaction strength is specified by the dimensionless quantity $u=U / \pi\Delta$.
Our analysis covers a wide parameter range from weak ($u=0.5$) to very strong ($u=5.0$) interactions
and moderate ($\beta=20$) to low ($\beta=2000$) temperatures. The corresponding Kondo temperatures are given in Tab.\ \ref{tab:kondo-temperatures}. (It should be noted that the two datasets for $\beta=20$ and $\beta=200$ have $u=0.5004$ rather than $u=0.5$. This minor difference changes the Kondo temperature by less than a factor $1.002$.)
To parametrize the vertex, different frequency conventions and index orderings can be used. Both are listed in \App{app:frequency_conventions}.
Finally, note that the spin structure of the vertex $\Gamma^{\sigma_1\sigma_2\sigma_3\sigma_4}$
can be simplified by exploiting the $\mathrm{SU}(2)$ spin symmetry of the SIAM in the absence of a magnetic field. Only components of the form
\begin{align}
    \Gamma^{\sigma\sigma'}=\Gamma^{\sigma\sigma\sigma'\sigma'}
\end{align}
are needed. Moreover, we have $\Gamma^{\downarrow\downarrow}=\Gamma^{\upup}$ and $\Gamma^{\downarrow\uparrow}=\Gamma^{\updown}$ by spin flip symmetry, such that only $\Gamma^{\upup}$ and $\Gamma^{\updown}$ remain independent.
The same applies to $\gamcore^{\sigma\sigma'}$.

\subsection{mpNRG vertex functions: Matsubara formalism}
\label{subsec:MatsubaraResults}
\begin{figure}[tbp]
\centering
\includegraphics[width=\columnwidth, trim=0.3cm 0 0.3cm 0, clip]{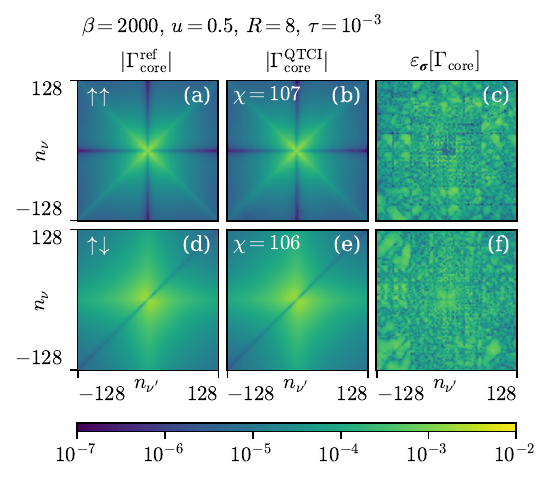}
\caption{
    QTCI-compression of the Matsubara core vertex $\gamcore(\omega,\nu,\nu')$ in the $p$-channel at $\beta=2000$, with $R=8$ and tolerance $\tcitoltxt 10^{-3}$. Heatmaps show the $\log_{10}$ absolute value of $\gamcore^{\upup}$ in (a,b) and $\gamcore^{\updown}$ in (d,e) on the slice $\omega=0$. 
    $n_\nu$ and $n_{\nu'}$ enumerate the fermionic Matsubara frequencies $\nu,\nu'$.
    Left column: Reference data $\gamcore^{\mathrm{ref}}$. Center column: QTCI representation $\gamcore^{\mathrm{QTCI}}$. Right column: Normalized error $\varepsilon_{\boldsymbol{\sigma}}[\gamcore]\lesssim1.58\cdot10^{-3}$ defined in \Eq{eq:tciconvergence}.
    We reproduce key features of the vertex on a large frequency box with a comparatively low QTT rank of $\chi=107$ and $\chi=106$, respectively.
}
\label{fig:MatsubaraVertexTriptych}
\end{figure}
Let us first consider the QTCI-compression of the Matsubara core and full vertices.
An important input to the QTCI algorithm is the specified error tolerance $\tau$, see \Eq{eq:tciconvergence}.
When compressing vertices from mpNRG, the choice of tolerance should be based on the accuracy of the 
PSFs. Based on benchmark results of Refs.\ \onlinecite{KuglerMultipointCorrelators,MultipointNRG,sIE2024}, we expect the mpNRG vertex to be reliable
to roughly $10^{-3}\!\cdot\!||\gamcore||_\infty$, where the error is partially systematic (as opposed to pure white noise). It should be emphasized that this
is only an estimate and inherent errors in \mbox{mpNRG} (due to discretization of the noninteracting bath and discarding high-energy eigenstates during iterative diagonalization) are different from those stemming from TCI.
A tolerance significantly below $\tcitoltxt 10^{-3}$ may be desirable for two reasons: First, to avoid errors \eqref{eq:tciconvergence} \textit{larger} than our mpNRG accuracy estimate of $10^{-3}$: After all, a local error \eqref{eq:tciconvergence} below the tolerance is only expected within the set of pivots that have been sampled by TCI
-- and even for these, the tolerance is not strictly guaranteed by the TCI routine used here (see \cite{Fernandez2024}, Sec.\ 4.3.1)
, which breaks full nesting conditions.
Lowering the tolerance increases the confidence that the required accuracy has been reached even outside the sampled set. The second motivation is to assess the potential of our approach for situations where more precise input data is available.
We shall therefore investigate tolerances ranging from $10^{-2}$ to $10^{-5}$.

The vertex functions \(\gamcore(\omega, \nu, \nu')\) and \(\Gamma(\omega, \nu, \nu')\) to be represented in QTT format here generally have prominent structures around the origin, along the frequency axes, and along the diagonals \cite{Rohringer2012}.
This is exemplified in \Fig{fig:MatsubaraVertexTriptych}, which
shows slices of $\gamcore$ at fixed bosonic frequency $\omega=0$.
The inverse temperature is \mbox{$\beta=2000$}.
We compare reference data with the
QTT representation of the vertex for a TCI tolerance of $\tcitoltxt 10^{-3}$. The reference was obtained by evaluating the vertex only on the two-dimensional slice shown in \Fig{fig:MatsubaraVertexTriptych}.
A logarithmic color scale has been chosen to expose imperfections of the TCI approximation.
\Fig{fig:MatsubaraVertexTriptych} illustrates how
QTCI represents important features of the vertex in a strongly compressed format:
For a $\threetimes{256}$ ($R=8$) frequency grid, we have ranks of
$\chi=107$ for $\gamcore^{\upup}$ and $\chi=106$ for $\gamcore^{\updown}$.
This translates to memory footprints reduced by factors of 92 (\memmb{268} to \memmb{2.9}) and
89 (\memmb{268} to \memmb{3.0}), respectively.

\begin{figure}[tbp]
    \includegraphics[width=\columnwidth, trim=0.3cm 0 0.3cm 0, clip]{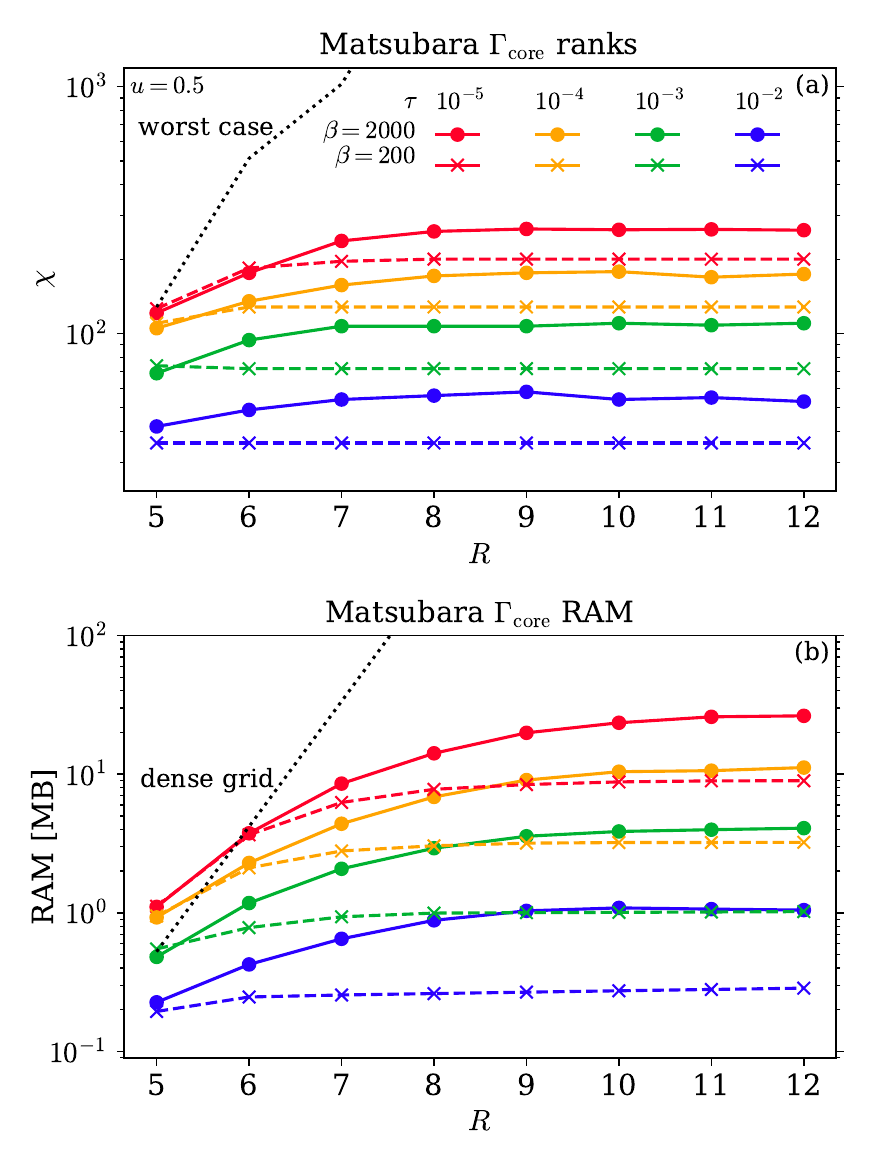}
    \caption{%
    (a) Rank and (b) RAM usage of the interleaved QTT representation of the Matsubara core vertex $\gamcore^{\upup}$ in the $p$-channel vs.\ frequency grid size for different tolerances.
    The grid has $2^R$ points in each frequency argument.
    For the target tolerance of $\tcitoltxt10^{-3}$, ranks saturate at $\chi \approx 100$.
    Dotted worst-case lines in (a) and (b) indicate the maximum rank of a $3R$-leg QTT (hence the even-odd alternation in the worst case of (a)) and the RAM
    requirements of dense grids with $2^{3R}$ points, respectively.
    }
    \label{fig:MatsubaraVertexRanks}
\end{figure}

\begin{figure*}[tbp]
\includegraphics[width=\textwidth, trim=0 0.5cm 0 0, clip]{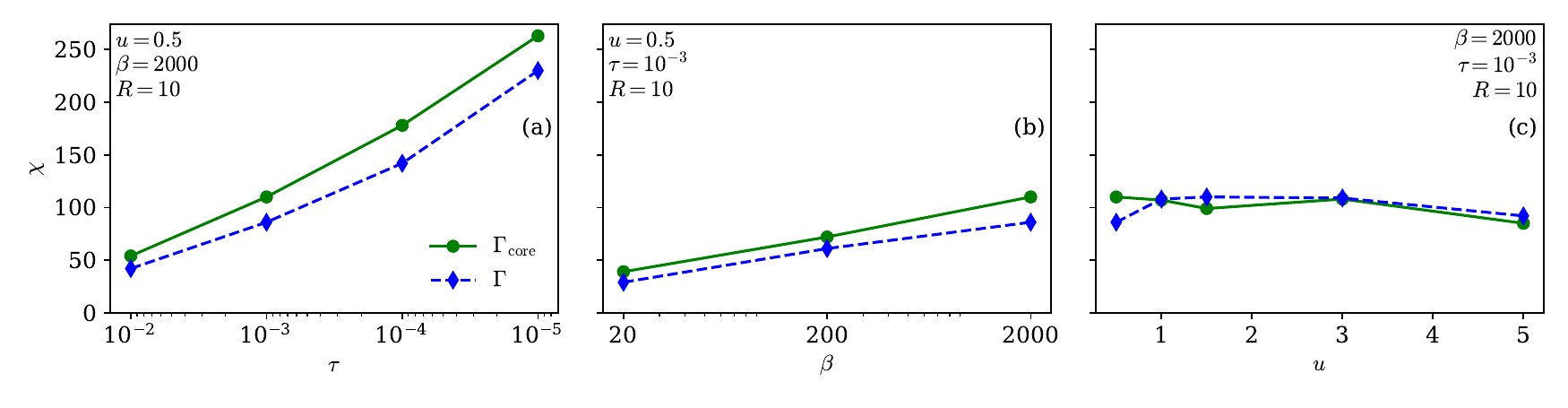}
\caption{%
Rank of the Matsubara core vertex $\gamcore^{\upup}$ and full vertex $\Gamma^{\upup}$ in the
$p$-channel versus (a) TCI tolerance $\tcitol$, (b) inverse temperature $\beta$ and (c) interaction
strength $u$.
}
\label{fig:MatsubaraParamsVSRank}
\end{figure*}

A systematic account of the compressibility of $\gamcore$ is provided in \Fig{fig:MatsubaraVertexRanks}.
It shows (a) ranks and (b) memory consumption of the resulting QTTs, as a function of tolerance $\tcitol$ and grid size.
The grid size is governed by the number $R$ of quantics bits in each dimension. 
Two different datasets at $\beta=200$ and $\beta=2000$ are represented by crosses and circles, respectively.
The QTT ranks do not exceed 250 (red circles, \mbox{$\tcitol=10^{-5}$}), with
compressions for the target accuracy of $\tcitoltxt10^{-3}$ (green) \textit{saturating} w.r.t.\ $R$ at $\chi=96$ even for the low temperature data. This rank saturation reflects the simple asymptotic structure of $\gamcore$.

In light of recent work by Rohshap et.~al.\ \cite{rohshap2024}, these results are very promising:
There, the authors demonstrate that self-consistent parquet calculations with maximum bond dimensions of 200 are feasible
on a single CPU (cf.\ Ref.\ \onlinecite{rohshap2024}, \mbox{Sec.~VI B}). While the calculations in Ref.~\onlinecite{rohshap2024} were performed
in a different parameter regime of the SIAM, their computational cost is determined by the bond dimensions of
the QTTs involved.
Our results therefore suggest that QTT-based parquet calculations with an NRG Matsubara vertex as input will be feasible.
\Fig{fig:MatsubaraVertexRanks} further shows that
TCI ranks of $\gamcore$ do not significantly increase beyond a grid size of $R=8$. In this region of saturated ranks,
both memory usage and runtime required for
manipulations of the vertex such as convolutions or frequency transformations scale logarithmically in the grid size (linearly in $R$) \cite{Ritter2024}.
In this regime, the QTT representation yields an exponential reduction in computational cost compared to dense grids. Lowering the tolerance (thus increasing $\chi$) comes at a runtime cost of $O(\chi^4)$ for the most expensive manipulations performed in
Ref.\ \onlinecite{rohshap2024} (see Sec.\ V.D there).

The linear scaling in $R$ generically allows for exponentially cheap reduction of discretization errors (for continuous variables) or errors due to finite-size domains (for discrete variables) \cite{Ritter2024}.
For Matsubara vertices, the asymptotic structure contains terms \(\gambare, \kclass_1, \kclass_2,\) and \(\kclass_{2'}\) that are independent of some of the frequencies (see Eq.~\eqref{eq:Kclasses}), implying that the function does not decay to zero at infinity \cite{KclassesToschi}.
One might be tempted to conclude that this makes any finite box representation invalid without high-frequency extrapolation.
In practice, vertex functions are used in evaluating diagrammatic equations such as expectation values of observables, Bethe--Salpeter equations or the Schwinger--Dyson equation. In all of these cases, the vertex is embedded in a frequency summation or integral with single-particle propagators that \textit{do} approach zero asymptotically for high frequencies.
The error generated in such summations and integrals is thus the relevant criterion for frequency box size, as was shown in Ref.~\onlinecite{rohshap2024} for a parquet approach, including Bethe--Salpeter equations.
There, the error of a QTCI-based self-consistent calculation of the density channel irreducible vertex was shown to improve from $10^{-1}$ for $R=5$ to $10^{-3}$ for $R=9$.
Frequency grids larger than $R=8$ are hence relevant.
We are considering significantly lower temperatures ($\beta D\in\{200,2000\}$, $U/\pi\Delta=0.5$ here vs.\ $\beta D=100$, $U/\pi\Delta\approx 0.51$ in Ref.\ \onlinecite{rohshap2024}).
Since $\gamcore$ becomes more complicated at these low temperatures (cf.\ \Fig{fig:MatsubaraParamsVSRank}(b) below), we expect
large grids to be even more relevant in NRG+parquet calculations.

To assess the range of applicability of our approach, we also examined how
the QTT rank for $\gamcore$ as well as the full vertex $\Gamma$ depends on the desired tolerance, inverse temperature $\beta$
and interaction strength $u$. The results for the $\upup$ flavor and an $R=10$ grid are summarized in \Fig{fig:MatsubaraParamsVSRank}.
The full and core vertices show a similar increase in rank with the TCI tolerance (\Fig{fig:MatsubaraParamsVSRank}(a)), since $\gamcore$ contains precisely the complex $3$-dimensional structure
of the full vertex.
Consistent with previous results on random pole based Matsubara correlators (cf.\ Fig.\ 8b in Ref.\ \onlinecite{RandomPole} ),
the TCI ranks increase with $\beta$, though only logarithmically, see \Fig{fig:MatsubaraParamsVSRank}(b).
Finally, \Fig{fig:MatsubaraParamsVSRank}(c) shows the ranks versus the interaction strength.
The key finding is that both vertices remain strongly compressible with ranks $\leq 110$ when
increasing $u$ from the
perturbative regime ($u\ll 1$) to very strong coupling ($u=5$). Since the $y$-axis ranges only from 85 to 110,
the observed variation in ranks with $u$ does not carry much significance.
Over all, \Fig{fig:MatsubaraParamsVSRank} suggests that parquet calculations with an mpNRG vertex as input will be feasible across a wide range of parameters.

\begin{figure*}[tp]
\includegraphics[width=\textwidth]{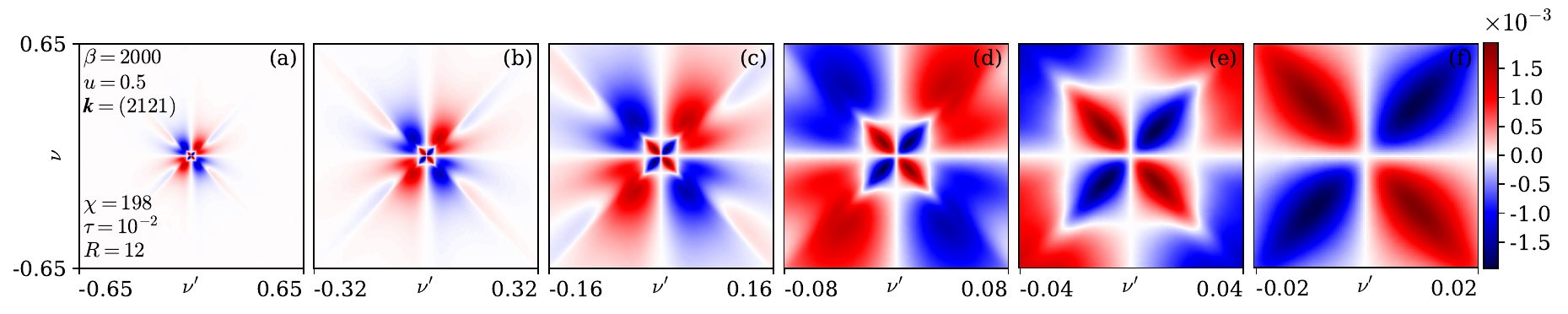}
\caption{Imaginary part of the Keldysh core vertex $\gamcore^{2121,\upup}$, at $\omega=0$,
compressed using $R=12$, $\tcitol=10^{-2}$. On this slice, $\mathrm{Re}(\gamcore^{2121,\upup})$ is a factor 25 smaller than $\mathrm{Im}(\gamcore^{2121,\upup})$. The QTT rank is $\chi=198$.
The QTT representation allows us to zoom in by a factor of $2^5$ (from left to right) while retaining
a sharp resolution throughout.
}
\label{fig:KeldyshZoom}
\end{figure*}

According to \Eq{eq:Kclasses}, the full vertex also contains lower-dimensional contributions
$K_1^r$ and $K_{2^{\bracketprime}}^r$. In \App{app:K1K2}, we verify that they have very small TCI ranks ($\chi\lesssim 20$ for $\tcitoltxt10^{-3}$)
compared to $\gamcore$ and $\Gamma$. Moreover, we focused on the $\upup$ component of the vertex in the $p$-channel.
In \App{app:channelflavorkeldysh}, we discuss how the ranks of $\gamcore$ and $\Gamma$ depend on the frequency channel and spin component.

\subsection{mpNRG vertex functions: Keldysh formalism}
\label{subsec:KeldyshResults}

\begin{figure}[tbp]
\includegraphics[trim=0.3cm 0cm 0.3cm 0cm, clip, width=\columnwidth]{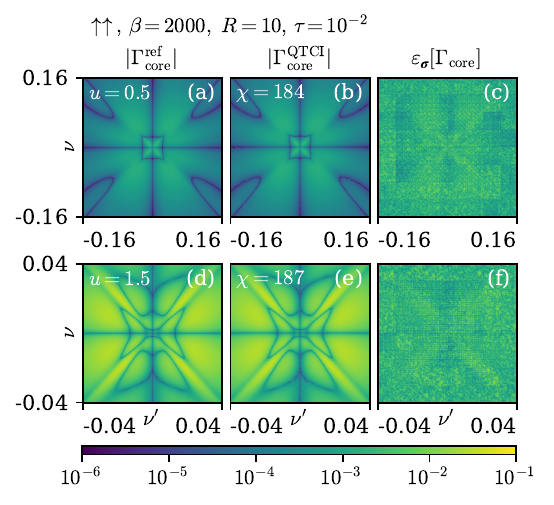}
\caption{%
QTCI-compression of the Keldysh core vertex \(\gamcore^{2121,\upup}(\omega, \nu, \nu')\) in the $p$-channel at $\beta=2000$, with $R=10$ and tolerance $\tcitol=10^{-2}$.
Heatmaps show the $\log_{10}$ absolute value of the vertex on the slice \(\omega = 0\).
The interaction strengths are $u=0.5$ in (a,b,c) and $u=1.5$ in (d,e,f).
Left column: Reference data $\gamcore^{\mathrm{ref}}$.
Center column: QTT representation $\gamcore^{\mathrm{QTCI}}$.
Right column: Normalized error $\varepsilon_{\boldsymbol{\sigma}}[\gamcore]\lesssim1.73\%$ defined in \Eq{eq:tciconvergence}.
The QTT representation captures complex features using moderate bond dimensions of $\chi=184$ and $\chi=187$, respectively.
}
\label{fig:KeldyshVertexTriptych}
\end{figure}

We now turn to computations of the Keldysh vertex in QTT format.
In contrast to the Matsubara vertex, this object gives direct access
to real-frequency dynamic response functions, but is a significantly more complicated function on a continuous domain of real frequencies.
Faithfully capturing its structure on a finite grid while keeping the computational cost in check is very challenging.
This has been achieved in Refs.\ \onlinecite{AnxiangNepomukKeldysh,NepomukAnxiangKeldyshCode}, but requires tedious manual tuning of nonlinear grids. The evaluation of mpNRG Keldysh vertices on nonlinear grids is discussed in \App{app:nonlin_V_KF}.
Our QTCI-based approach allows us to automatically capture features on different length
scales on an extremely fine equidistant grid.
Our grid for $\omega$ contains 0, while $\nu$ and $\nu'$ live on a grid that is offset from 0 by half a grid spacing. An alternative choice would be to include 0 in all three grids. In QTCI, we can refine the grid until all features are represented up to a given tolerance, so that shifting the $\nu$ and $\nu'$ grids by half a grid spacing does not make a difference. The resolution attained with QTCI is exemplified in \Fig{fig:KeldyshZoom}, showing a QTT representation
of $\gamcore^{2121,\upup}$ on a slice at $\omega=0$ using \(R = 13\) quantics bits.
The TCI tolerance was set to $10^{-2}$.
All panels show the same slice, but zoom in by factors of 2 moving from left to right.
The rightmost panel still exhibits a sharp resolution after a 32-fold magnification.
As a further illustration, \Fig{fig:KeldyshVertexTriptych} compares the TCI-compressed
vertex (center) to the reference (left), showing the normalized error on the right.
The slices are taken again at $\omega=0$ and for $u=0.5$ (top row) and $u=1.5$ (bottom row).
Overall, we see that TCI resolves the core vertex to $1\%$ precision with ranks of $184$ and $187$, respectively.

This $1\%$ error is comparable to the uncertainty due to the broadening of spectral functions, which supersedes the NRG error of $10^{-3}$ in the real-frequency case:
While there are well-established schemes \cite{MultipointNRG,sIE2024} to choose broadening parameters, 
legitimate choices can vary within a range that causes vertex functions to change by a few percent.
Our default tolerance for Keldysh objects is therefore chosen as $\tcitoltxt 10^{-2}$.
In view of ongoing research aiming to develop an impurity solver less susceptible to broadening artifacts \cite{Picoli2025}, which could be extended to the multipoint case in the future, we extend our investigations
down to $\tcitoltxt 10^{-3}$.

Figure \ref{fig:KeldyshVertex_tolvsrank} shows (a) the rank and (b) the memory size of the compressed Keldysh core vertex component $\gamcore^{2121,\upup}$ at temperatures \(\beta = 200\) and \(2000\) versus the number of
quantics bits $R$ in each dimension. This Keldysh component of $\gamcore^{\boldsymbol{k}}$ was found to have the highest bond dimension (see \App{app:channelflavorkeldysh}, \Fig{fig:RanksvsiK}).
We set a fixed box size of $\omega_{\max}=0.65$ 
(cf.\ \Fig{fig:KeldyshZoom}(a)) and increase the resolution with $R$.
We verified that the chosen box size is large enough to capture all relevant structures within the target tolerance $\tcitoltxt10^{-2}$.
At this tolerance, the rank shown in \Fig{fig:KeldyshVertex_tolvsrank} saturates at \(\chi = 202\). This rank is again of a magnitude where a self-consistent parquet calculation in the Matsubara formalism was shown to be feasible on a single core in Ref.\ \onlinecite{rohshap2024}. The 16 components of the Keldysh vertex can be inferred from just 5 components using complex conjugation and crossing symmetry \cite{Jakobs2010}. Nevertheless, in follow-up computations such as solving the parquet equations, these multiple Keldysh components in contrast to a single Matsubara vertex may necessitate parallelization already for $\chi\approx200$.
Multithreaded or distributed schemes will certainly be required for the most difficult case considered here (\(\tau = 10^{-3}\) and \(\beta = 2000\)), which results in ranks of \(\chi\approx450\).
On a different note, the QTT vertex has a vastly reduced memory footprint, as shown in \Fig{fig:KeldyshVertex_tolvsrank}(b): For $\beta=2000$, $\tcitoltxt10^{-2}$ and $R=10$, it requires \(11.3~\mathrm{MB}\) of memory, compared to \(17.1~\mathrm{GB}\) for a dense grid representation; this corresponds to a compression ratio of \(1:1513\).
Although real-frequency diagrammatic calculations for the SIAM are limited by runtime rather than memory \cite{AnxiangNepomukKeldysh}, this paves the way for investigation of more complicated models with multiple orbitals or momentum dependence, which have prohibitive memory requirements if attempted with dense grids \cite{Eckhardt2020,Krien2022a,Li2019}.

\begin{figure}[tbp]
\includegraphics[width=\columnwidth, trim=0.4cm 0 0.3cm 0, clip]{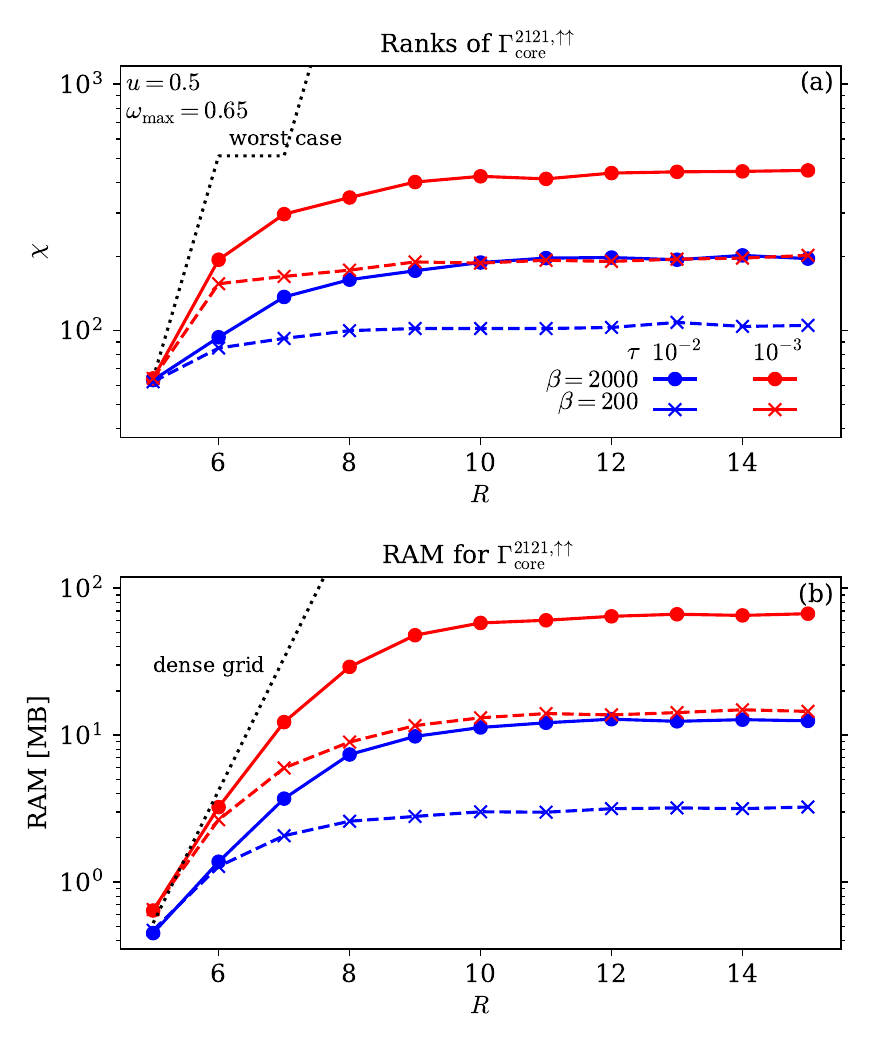}
\caption{%
(a) Rank and (b) RAM usage of Keldysh core vertex $\gamcore^{2121,\upup}(\omega, \nu, \nu')$ in the $p$-channel vs.\ frequency
grid size for different tolerances and temperatures. The grid has $2^R$ points in each frequency argument.
For a tolerance of $\tcitol=10^{-2}$ (blue), the bond dimension saturates at \(\chi \approx 200\).
Dotted worst-case lines in (a) and (b) indicate the maximum rank of an $R$-leg QTT (fused representation) and the RAM
requirements of dense grids with $2^{3R}$ points, respectively.
}
\label{fig:KeldyshVertex_tolvsrank}
\end{figure}

In \Fig{fig:KeldyshParamsVSRank} we explore the compressibility of the core and full vertices for varying
tolerance \(\tau\), inverse temperature \(\beta\) and interaction strength \(u\).
As seen before in \Fig{fig:KeldyshVertex_tolvsrank}, lowering the tolerance below $10^{-2}$ results in
a steep increase in the rank. As in the Matsubara formalism, the rank increases with $\beta$, but only
slowly.
Panel (c) reveals a much less predictable behavior: The ranks of both the full and core vertices
reach a maximum at $u=1.0$ and decrease significantly for large $u$.
Moreover, the rank of the full vertex $\Gamma$ approaches that of $\gamcore$ from below with increasing interaction $u$.
This reflects the increasing magnitude of $\gamcore$ relative to the asymptotic contributions $\kclass_1^r$ and $\kclass_{2^{\bracketprime}}^r$: At weak interaction, the magnitude of the core vertex is much smaller than that of the full vertex, which is dominated by $\kclass_1^r$ and $\kclass_{2^{\bracketprime}}^r$. Since TCI measures the error relative to the supremum norm of the target function (cf.\ \Eq{eq:tciconvergence}, this means that $\gamcore$ need not be resolved as accurately at weak interaction.
The compression of the asymptotic contributions $\kclass_1^r$ and $\kclass_{2^{\bracketprime}}^r$ is discussed in \App{app:K1K2}, together with their Matsubara counterparts.
In \App{app:channelflavorkeldysh}, we discuss how the ranks of $\gamcore^{\boldsymbol{k}}$ and $\Gamma^{\boldsymbol{k}}$ depend on flavor, frequency channel
and Keldysh component $\boldsymbol{k}$.

Finally, we discuss how compressing each Keldysh component of the vertex separately, as was done in this work, compares to running TCI on the entire Keldysh core or full vertex, where the Keldysh components are encoded in an additional leg of a single tensor train. In both cases, spin components are compressed separately. For a fair comparison of these two approaches, recall the following: (i) TCI measures the interpolation error relative to the supremum norm of the target function (see \Eq{eq:tciconvergence}). When compressing the entire vertex, any given Keldysh component $\Gamma^{\boldsymbol{k}}$ (or $\gamcore^{\boldsymbol{k}}$) should therefore be normalized by $||\Gamma^{\boldsymbol{k}}||_{\infty}$ (or $||\gamcore^{\boldsymbol{k}}||_{\infty}$), i.e., with the supremum norm of the same Keldysh component $\boldsymbol{k}$. Only then does one achieve the same accuracy as in separate compressions of Keldysh components. (ii) MPO--MPO contractions, the most expensive operations in QTT-based diagrammatic calculations, scale as $\order{R\chi^4}$ in runtime.
As a preliminary investigation, we compressed the entire core vertex at $u=0.5,\beta=2000$ and $\omega_{\max}=0.65$ with a tolerance of $\tcitoltxt10^{-2}$ and $R=8$ quantics bits. The tensor leg for the Keldysh component was placed to the very left and only included the five Keldysh components not related by crossing symmetry or complex conjugation (cf.\ \App{app:channelflavorkeldysh}).
\begin{figure*}[tbp]
\includegraphics[width=\textwidth, trim=0 0.5cm 0 0, clip]{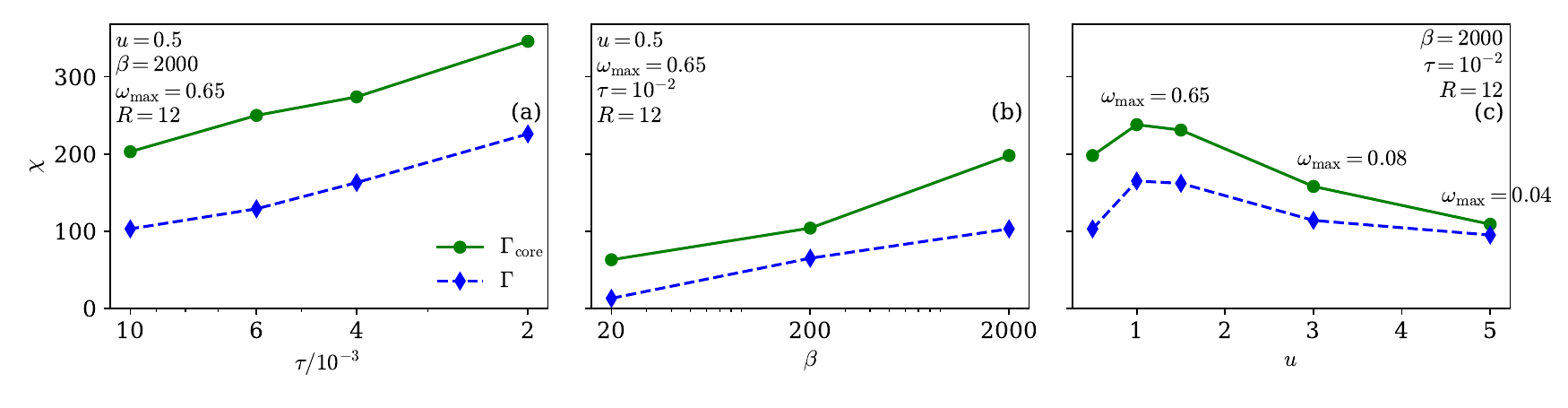}
\caption{Rank of Keldysh core vertex $\gamcore^{\upup}$ and full vertex $\Gamma^{\upup}$ in the $p$-channel versus
(a) TCI tolerance $\tcitol$, (b) inverse temperature $\beta$ and (c) interaction strength $u$. We show data for the $\boldsymbol{k}=(2121)$ Keldysh component. In panel (c), we choose smaller box sizes $\omega_{\max}=0.08$ and $\omega_{\max}=0.04$ for $u=3.0$ and $u=5.0$, respectively. This is because the extent of the core vertex decreases at these strong interactions.}
\label{fig:KeldyshParamsVSRank}
\end{figure*}
\begin{table}[t]
    \centering
    \begin{tabular}{c|cccccc}
        component $\boldsymbol{k}$ & 1111 & 2111 & 2121 & 2112 & 1222 & \textbf{all}\\
         \midrule
        rank $\chi$ & 130 & 136 & 159 & 126 & 95 & \textbf{344}
    \end{tabular}
    \caption{QTT ranks of the five Keldysh components not related by crossing symmetry or complex conjugation (center columns) compared to the QTT rank of a single 
    tensor that contains all five components (rightmost column). Parameters are $u=0.5,\beta=2000,\omega_{\max}=0.65,\tau=10^{-2}$ and $R=8$.}
    \label{tab:Keldyshranks}
\end{table}
The resulting rank $\chi$ is compared with the ranks $\chi_{\boldsymbol{k}}$ of individual Keldysh components in Tab.\ \ref{tab:Keldyshranks}.
We observe the ratio $\chi^4/\max_{\boldsymbol{k}}\chi_{\boldsymbol{k}}^4\approx21.77$, 
thus MPO-MPO contractions take about 22 times longer for two entire vertices than for two individual components. On the 
other hand, the latter type of contraction
would have to be performed $5^2 = 25$ times. However, Tab.\ \ref{tab:Keldyshranks} shows that some Keldysh components have a significantly lower bond dimension than $\max_{\boldsymbol{k}}\chi_{\boldsymbol{k}}$. In summary, both approaches are worth investigating, and a conclusive comparison is only possible in the context of a specific QTT-based diagrammatic code.}

\section{Summary and Outlook}
\label{sec:SummaryOutlook}
We have presented a QTCI-based method for representing imaginary- and real-frequency mpNRG vertex
functions on large grids that are far beyond the reach of previous implementations.
The QTCI algorithm allows us to automatically capture all relevant features of the vertex up to a prescribed
accuracy and represents the result in a highly compressed format. Repeated sampling during TCI sweeps
necessitates optimizations of the vertex evaluation, which we described in detail. We studied the compressibility of
the vertex in a systematic fashion: 
Imaginary- and even real-frequency vertices
are representable as QTTs with maximum bond dimensions sufficiently small ($\chi\approx$ a few hundred) to allow for diagrammatic computations with these
objects.
This holds true across a broad range of temperatures and interaction strengths, both
for the full vertex as well as its asymptotic decomposition.
Our work thus constitutes an important step toward QTCI-based diagrammatic calculations which use a nonperturbative DMFT vertex as input,
and suggests that these will be feasible.
The next step will be to implement a QTCI-based diagrammatic extension   
of DMFT that augments the local vertex with momentum dependence.
An analogous program can be envisioned in the real-frequency setting.
Though this is a challenging, computationally demanding endeavor, it would
achieve a long-sought goal: a method to obtain nonlocal, real-frequency dynamical response functions of strongly correlated systems.

\section*{Data and code availability}
The mpNRG computations were performed with the MuNRG package \cite{MuNRG2,MuNRG3,MultipointNRG}, which is based on the QSpace tensor network library \cite{Weichselbaum2012,WeichselbaumNRG2012,Weichselbaum2020,Weichselbaum2024}.
The latest version of QSpace is available \cite{QSpace4} and a public release of MuNRG is intended. The code used in this work to compute and compress vertices is available on GitHub, see Ref.\ \onlinecite{TCI4Keldysh}. The partial spectral functions required as input for that code can be found in Ref.\ \onlinecite{data}.

\begin{acknowledgments}
We thank Seung-Sup Lee and Jae-Mo Lihm for helpful discussions.
We thank Seung-Sup Lee for providing the mpNRG code used for computing the PSFs.

This work was funded in part by the Deutsche Forschungsgemeinschaft under Germany's Excellence Strategy EXC-2111 (Project No.\ 390814868).
It is part of the  Munich Quantum Valley, supported by the Bavarian state government with funds from the Hightech Agenda Bayern Plus.
We further acknowledge support from the DFG grant LE 3883/2-2.
We gratefully acknowledge computational resources from grant INST 86/1885-1 FUGG of the German
Research Foundation (DFG) and from the GCS Supercomputer SuperMUC-NG at the Leibniz Supercomputing Centre in Munich provided by the Gauss Centre for Supercomputing e.V.
MF, AG, MR, and NR acknowledge support from the International Max-Planck Research School for Quantum Science and Technology (IMPRS-QST). NR acknowledges funding from the Studienstiftung des deutschen Volkes and the Marianne-Plehn-Programm of the state of Bavaria.
\end{acknowledgments}

\appendix

\section{Computing broadened Keldysh kernels}
\label{app:keldysh_broadening}
In this section, we provide details on the numerical computation of the broadened Keldysh kernels $\bretkernel{\lambda}$ appearing in \Eq{eq:retardedkernel}.
It consists of two steps: broadening the Dirac-$\delta$ functions in \Eq{eq:PSFdelta} to $\delta_b(\omega',\epsilon)$, and convolution
of $\delta_b(\omega',\epsilon)$ with the Keldysh kernel $(\omega'+\mi\imagshift)^{-1}$. The numerical details of this procedure are taken from the MuNRG code of Ref.\ \onlinecite{MultipointNRG}.

The broadening combines symmetric log-Gaussian and linear broadening (cf.\ Ref.\ \onlinecite{sIE2024}, App.\ E.2):
\begin{subequations}
\label{eq:broadening}
\begin{align}
    \label{eq:app_deltab}
    \delta_b(\omega',\epsilon)&=\int_{\mathbb{R}}\md\epsilon'\delta_{\mathrm{F}}(\omega',\epsilon')\delta_{\mathrm{sLG}}(\epsilon',\epsilon),\\
    \delta_{\mathrm{sLG}}(\epsilon',\epsilon)&=\frac{\Theta(\epsilon'\epsilon)}{\sqrt{\pi}\sigma_{\mathrm{sLG}}|\epsilon|}
        \exp\left[-\left(\frac{\ln|\epsilon/\epsilon'|}{2\sigma_{\mathrm{sLG}}} - \frac{\sigma_{\mathrm{sLG}}}{4}\right)^2\right],\\
    \delta_{\mathrm{F}}(\omega',\epsilon')&=\frac{1}{2\gamma_\mathrm{L}}\left(1+\cosh\frac{\omega'-\epsilon'}{\gamma_{\mathrm{L}}}\right)^{-1}.
\end{align}
\end{subequations}
The broadening parameters $\gamma_{\mathrm{L}}$ and $\sigma_{\mathrm{sLG}}$ along with other numerical settings for
all physical parameter sets are listed in Tab.\ \ref{tab:broadening}. However, the linear broadening $\gamma_L$ is multiplied with a prefactor that depends on the current permutation $p$, the fully retarded index $\lambda$ and the dimension $i$. This scheme will be explained further at the end of this section (see \Eq{eq:varyinggamma}).
The integral \eqref{eq:app_deltab} is performed by trapezoidal quadrature, where
$\omega'$ and $\epsilon'$ are discretized on logarithmic grids. These grids contain $0$, are symmetric
around the origin and range from $e_{\min}$ to $e_{\max}$ with $e_{\mathrm{step}}$ points per decade (see Tab.\ \ref{tab:broadening}).
For the $\epsilon'$ grid, $e_{\min}$ is automatically replaced by a lower boundary $x_{\min}$ with $0<x_{\min}<e_{\min}$ if the low-frequency tail of the log-Gaussian broadening kernel
extends below $e_{\min}$. This ensures an accurate integration of $\delta_{\mathrm{sLG}}(\epsilon',\epsilon)$. 
The energies $\epsilon$ specifying the location of the spectral function peaks also reside on a logarithmic grid,
which arises from the mpNRG computation.

The numerical integration described above yields $\delta_b(\omega',\epsilon)$ with $\omega'$ and $\epsilon$ on logarithmic grids.
To convolve $\delta_b$ with the Keldysh kernel $(\omega'+\mi\imagshift)^{-1}$, we use the identity
\begin{equation}
\begin{aligned}
    \label{eq:app_KramersKronig}
    &\lim_{\gamma_0\rightarrow 0^+}\int_{\mathbb{R}}\md\omega'\frac{\delta_b(\omega',\epsilon)}{\omega-\omega'+\mi\imagshift}=\\
        &=\mathcal{P}\int_{\mathbb{R}}\md\omega' \frac{\delta_b(\omega',\epsilon)}{\omega-\omega'} - \mi\pi\mathrm{sgn}(\imagshift)\delta_b(\omega,\epsilon),
\end{aligned}
\end{equation}
where $\mathcal{P}$ denotes the Cauchy principal value (PV) integral. Also, recall
the definition \eqref{eq:imagshift} of $\imagshift$.
Importantly, $\delta_b(\omega',\epsilon)$ has been computed on a logarithmic grid, while $\omega$ in the broadened Keldysh kernel $\bretkernel{\lambda}(\omega,\epsilon)$ defined in \Eq{eq:retardedkernel} resides on a linear grid.
This is because the external frequency grids on which we compute vertices are also linear.
To obtain the imaginary part of \Eq{eq:app_KramersKronig} on the linear grid, we use linear interpolation of $\delta_b(\omega,\epsilon)$ in the argument $\omega$.
Computing the real part, i.e., the PV integral is slightly more involved: 
By the linear interpolation performed for the imaginary part, $\delta_b(\omega,\epsilon)$ can be viewed
as a piecewise linear function. We split the PV integral over $\delta_b(\omega',\epsilon)$ into
PV integrals over linear functions $(a_{i,\epsilon}\omega'+b_{i,\epsilon})$ on intervals $[\omega_i',\omega_{i+1}']$.
These are evaluated using the formula
\begin{equation}
    \label{eq:PVlinear}
    \begin{aligned}
    &\mathcal{P}\int_{\omega_i'}^{\omega_{i+1}'}\!\!\!\md\omega'\frac{a_{i,\epsilon}(\omega'-\omega_i)+b_{i,\epsilon}}{\omega-\omega'}=\\
    &-a_{i,\epsilon}(\omega_{i+1}'-\omega_i') - \left(a_{i,\epsilon}(\omega-\omega_i') + b_{i,\epsilon}\right) \ln\bigg|\frac{\omega-\omega_{i+1}'}{\omega-\omega_i'}\bigg|.
    \end{aligned}
\end{equation}
The sum over all PV integrals of the form \eqref{eq:PVlinear} then yields the real part of the broadened Keldysh kernel $K_b(\omega,\epsilon)$.
For this scheme to be accurate, the extent $[-e_{\max},e_{\max}]$ of the logarithmic $\omega'$ grid should be significantly larger
than the frequency box delimited by $\omega_{\max}$. Comparing the values of $e_{\max}$ given in Tab.\ \ref{tab:broadening} with our
default frequency box size $\omega_{\max}=0.65$, one verifies that this is the case.
\begin{table}[tbp]
    \setlength{\tabcolsep}{5pt}
    \centering
    \begin{tabular}{cc|cc}
       $u$ & $\beta$ & $\sigma_{\mathrm{sLG}}$ & $\gamma_{\mathrm{L}}$\\
       \midrule
       $0.5$ & $20/200$ & $0.693$ & $T$\\
       $0.5/1.0/1.5$ & $2000$ & $0.4$ & $T$\\
       $3.0/5.0$ & $2000$ & $0.4$ & $T$\\
    \end{tabular}
    \caption{Broadening settings for different NRG datasets. $T=1/\beta$ denotes the temperature.
    See main text for definitions of the parameters.
    We set $e_{\min}=10^{-6},e_{\max}=10^4$ and $e_{\mathrm{step}}=50$. For TCI tolerances $\tau\leq 3.4\cdot 10^{-3}$, the integration grid was refined to $e_{\mathrm{step}}=200$ to avoid fitting of numerical noise by the TCI algorithm. We also set $e_{\mathrm{step}}=200$ to broaden 2p functions.}
    \label{tab:broadening}
\end{table}

We now turn to the prefactors of the linear broadening $\gamma_{\mathrm{L}}$ mentioned above.
In the linear broadening kernel $\delta_{\mathrm{F}}(\omint{i}',\epsilon_i')$ to be convolved with the Keldysh kernel $(\omint{i}'+\imagshift)^{-1}$, the broadening with $\gamma_{\mathrm{L}}$ is replaced by
\begin{equation}
\label{eq:varyinggamma}
\gamma_{\mathrm{L},i}^{\lambda}=\begin{cases}
        \gamma_{\mathrm{L}}\cdot (\ell-i)\quad\mathrm{for}\quad i\geq\lambda,\\
         \gamma_{\mathrm{L}}\cdot i\quad\mathrm{for}\quad i<\lambda.\\ 
    \end{cases}
\end{equation}
This choice was found to reduce broadening artifacts in an mpNRG treatment of the Hubbard atom in Ref.\ \onlinecite{JaeMoMail}. Moreover, composite operators $q_{ij}$ in 3p correlators (cf.\ Ref.\ \onlinecite{sIE2024}, Eq.\ (96)) receive a doubled linear broadening. This was found to cancel discretization and broadening artifacts when computing $\kclass_{2^{\bracketprime}}^r$ by multiplication with self-energies (see \Eq{eq:K2t}), see Ref.\ \onlinecite{JaeMoMail}. The broadening of 3p correlators is exemplified in Tab.\ \ref{tab:3ptbroaden}.

\begin{figure*}[t]
\includegraphics[width=\textwidth]{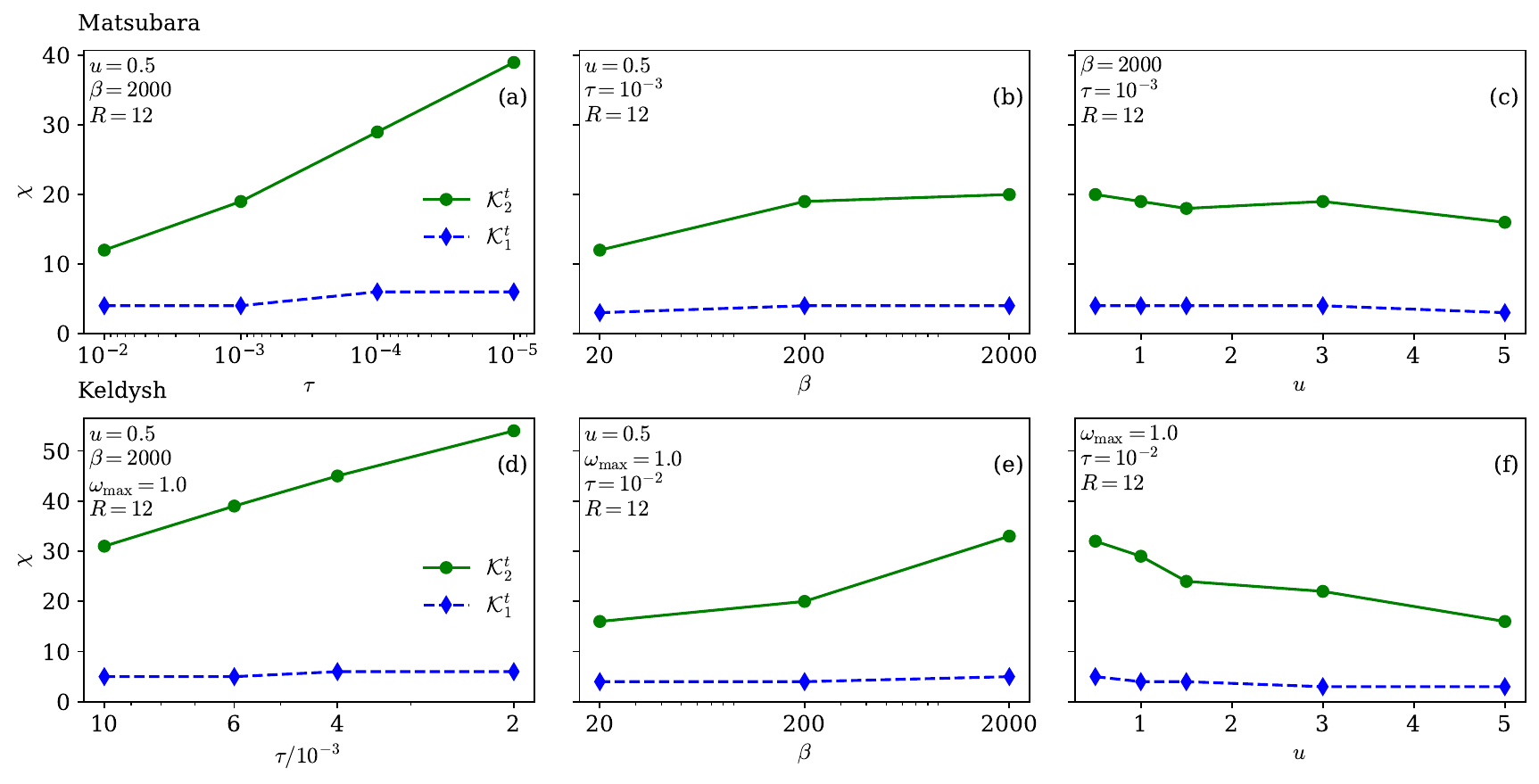}
\caption{%
QTT ranks of Matsubara (top row, (a--c)) and Keldysh (bottom row, (d--f)) $\kclass_1^{t,\upup}$ and $\kclass_2^{t,\upup}$
contributions to the full vertex versus: (a,d) tolerance $\tcitol$, (b,e) inverse temperature $\beta$ and (c,f) interaction strength $u$.
In Keldysh we chose, of all components, the component $\boldsymbol{k}=(22)$ for $\kclass_1^{t,\upup}$
and $\boldsymbol{k}=(112)$ for $\kclass_2^{t,\upup}$. These components were found to have the highest rank, respectively.}
\label{fig:K1K2}
\end{figure*}

\begin{table}[tbp]
        \centering
        \begin{minipage}{0.49\columnwidth}
            \centering
            \begin{tabular}{c|c|c}
                  & $i=1$ & $i=2$ \\
                \midrule
                $\lambda=1$ & $2\gamma_{\mathrm{L}}$ &  $\gamma_{\mathrm{L}}$\\
                $\lambda=2$ & $2\gamma_{\mathrm{L}}$ &  $\gamma_{\mathrm{L}}$\\
                $\lambda=3$ & $2\gamma_{\mathrm{L}}$ &  $3\gamma_{\mathrm{L}}$\\
            \end{tabular}
        \end{minipage}
        \hfill
        \begin{minipage}{0.49\columnwidth}
            \centering
            \begin{tabular}{c|c|c}
                & $i=1$ & $i=2$ \\
                \midrule
                $\lambda=1$ & $3\gamma_{\mathrm{L}}$ &  $\gamma_{\mathrm{L}}$\\
                $\lambda=2$ & $\gamma_{\mathrm{L}}$ &  $\gamma_{\mathrm{L}}$\\
                $\lambda=3$ & $\gamma_{\mathrm{L}}$ &  $3\gamma_{\mathrm{L}}$\\
            \end{tabular}
        \end{minipage}
        \caption{Linear broadening of a 3p correlator with doubled broadening on the composite operator $q_{ij}$. The operator $q_{ij}$ is in the first slot of the operator tuple for the identity permuation $p=[123]$.
        Left: Permutation $p=[123]$. Right: Permutation $p=[213]$.}
        \label{tab:3ptbroaden}
\end{table}
Finally, the 2p correlators required for self-energies in the symmetric estimators for $\gamcore$ and $K_{2^{\bracketprime}}^r$ (cf.\ \Sec{subsec:sIE} and \App{app:K1K2}.) are broadened according to Tab.\ \ref{tab:SEbroaden}.
 \begin{table}[tpb]
        \centering
        \begin{tabular}{c|c|c}
          & $i=1$, $p=[12]$  & $i=1$, $p=[21]$\\
            \midrule
            $\lambda=1$ & $3\gamma_{\mathrm{L}}$ & $\gamma_{\mathrm{L}}$\\
            $\lambda=2$ & $\gamma_{\mathrm{L}}$ & $3\gamma_{\mathrm{L}}$\\
        \end{tabular}
        \caption{Linear broadening of a 2p function used for the aIE self-energy. The two rightmost columns correspond
        to the two possible permutations.}
        \label{tab:SEbroaden}
\end{table}

\begin{figure}[tbp]
    \includegraphics[width=\columnwidth]{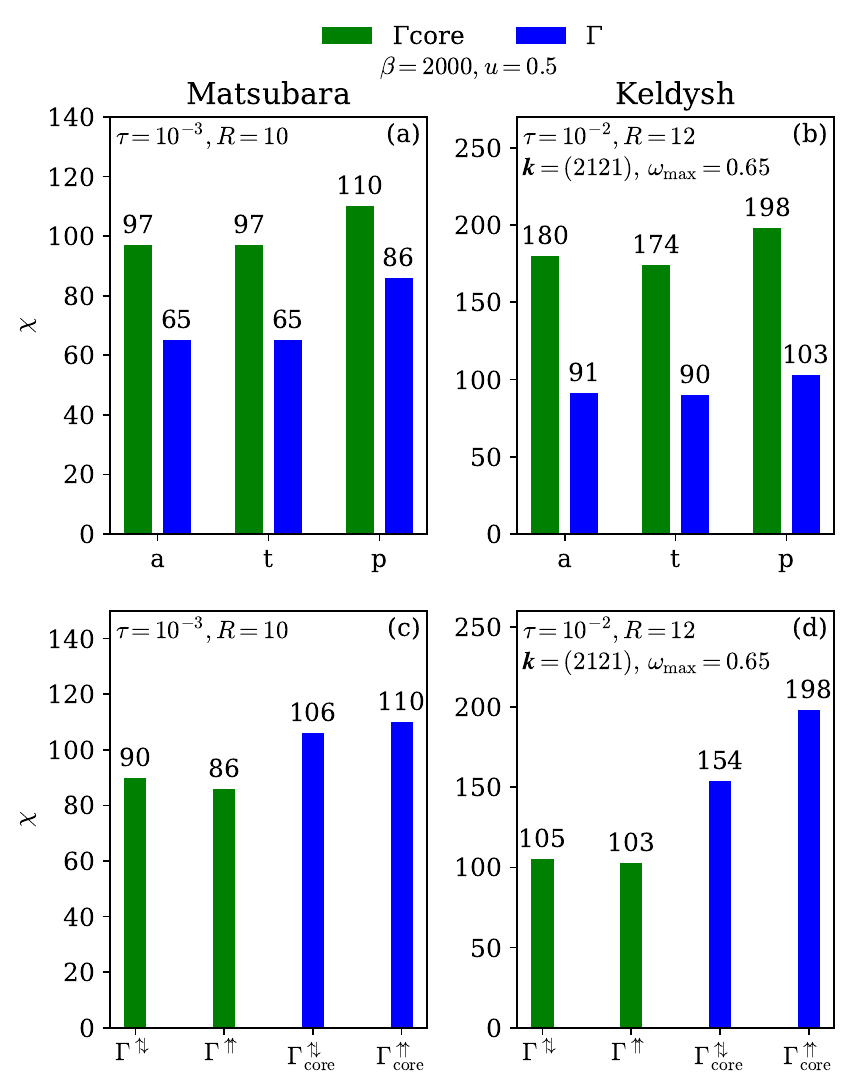}
    \caption{
Top row (a,b): TCI ranks of $\Gamma^{\upup}$ and $\gamcore^{\upup}$ in the three channels $a,p,t$.
Bottom row (c,d): TCI ranks of $\Gamma$ and $\gamcore$ in the $p$-channel for the two flavors $\upup$ and $\updown$.
Matsubara vertices (a,c) were compressed with
$\tcitol=10^{-3}$ and $R=10$, Keldysh vertices (b,d) with $\tcitol=10^{-2}$ and $R=12$.
}
\label{fig:ranksvsflavorchannel}
\end{figure}

\begin{figure}[tbp]
    \includegraphics[width=\columnwidth, trim=0.5cm 0.5cm 0 0, clip]{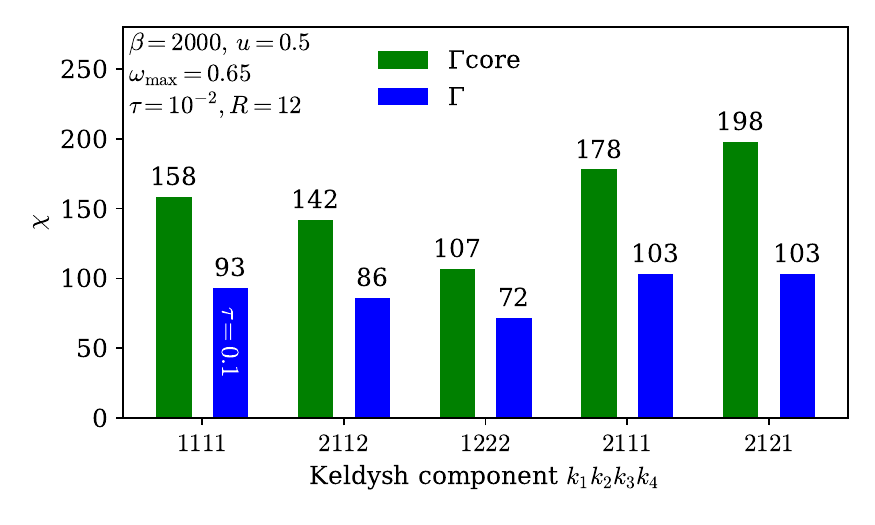}
    \caption{QTT rank of Keldysh core vertex $\gamcore^{\boldsymbol{k},\upup}$ in the $p$-channel
    vs.\ Keldysh component. From the 16 Keldysh components, only those are shown which
    are not related by crossing symmetry or complex conjugation.
    The $\Gamma^{1111,\upup}$ component of the full vertex was compressed with tolerance $\tau=0.1$, since it is about a factor 10 smaller than the other Keldysh components of the full vertex. This is because the $\kclass_1^r$ contributions, which dominate other components $\Gamma^{\boldsymbol{k}\neq1111}$, vanish in the $\boldsymbol{k}=(1111)$ component.
    }
\label{fig:RanksvsiK}
\end{figure}

\section{Compression of 1D and 2D vertex contributions}
\label{app:K1K2}
A QTCI-based parquet calculation exploiting the asymptotic decomposition \eqref{eq:Kclasses} requires not only $\gamcore$,
but also $\kclass_1^r$ and $\kclass_{2^{\bracketprime}}^r$ represented as QTTs.
Recall that $r=a,p,t$ labels the three frequency channels.
In this section, we verify that these asymptotic contributions indeed have a significantly lower rank than $\gamcore$, as expected from their
simpler structure.

The $\kclass_1^r$ contributions are simply given by 2p correlators of composite operators, see Ref.\ \onlinecite{sIE2024}, Sec.\ IV.F.
We illustrate the evaluation of $\kclass_{2^{\bracketprime}}^r$ using $\kclass_2^t$ as an example. For derivations and the remaining $\kclass_{2^{\bracketprime}}^r$ components we refer to Ref.~\onlinecite{sIE2024}, Secs.\ IV.C and IV.F.
Since the self-energy is spin-diagonal, we again omit spin indices.
First the operator $q=[d,H_{\mathrm{int}}]$ introduced in \Sec{subsec:sIE} is used to define the operator $q_{34}=\{q,d^{\dag}\}=qd^{\dag}+d^{\dag}q$.
One further introduces 3p correlators $G[q_{34},a_1,a_2^{\dag}]$, where $a_1,a_2\in\{d,q\}$.
They are defined in terms of connected correlators as
\begin{equation}
         G^{\boldsymbol{k}}[q_{34},a_1,a_2^{\dag}]=P^{k_1k_2(k_3+k_4)}G^{\boldsymbol{k}}_{\mathrm{con}}[q_{34},a_1,a_2^{\dag}].
\end{equation}
In the Keldysh formalism, the tensor $P$ reads
\begin{equation}
    P^{k_1k_2(k_3+k_4)}=
    \begin{cases}
        \frac{1}{\sqrt{2}}\quad &\text{if } \sum_i k_i \text{ is odd},\\
        \hfil 0 \quad &\text{else},\\
    \end{cases}
\end{equation}
while it is set to unity in Matsubara.
Using the symbol $Y_{x_i}$ introduced in \Eq{eq:Ysymbol}, $K_2^t$ can then be expressed as:
\begin{equation}
    \label{eq:K2t}
    \kclass_2^t(\omega_t,\nu_t) = \sum_{a_1,a_2\in\{d,q\}}Y_{a_1}G[q_{34},a_1,a_2^{\dag}](-\omega_{12},\omega_1,\omega_2)Y_{a_2}.
\end{equation}
To evaluate $\kclass_2^t$, the external frequencies $\omega_1,\omega_2$ appearing on the RHS of \Eq{eq:K2t} are expressed
in the $t$-channel parametrization, i.e., in terms of $\omega_t$ and $\nu_t$.

The ranks of Matsubara and Keldysh asymptotic contributions $\kclass_{1}^{t,\upup}$, $\kclass_{2}^{t,\upup}$
for different parameters are shown in \Fig{fig:K1K2}(a--c) and \ref{fig:K1K2}(d-f), respectively.
We use the $t$-channel frequency parametrization, thus viewing $\kclass_1^t(\omega_t)$ as a 1D and $\kclass_2^t(\omega_t,\nu_t)$ as a 2D function.
A comparison of \Fig{fig:K1K2} with Figs.\ \ref{fig:MatsubaraParamsVSRank} and \ref{fig:KeldyshParamsVSRank}
confirms that the three-dimensional vertex functions will dominate the cost of a diagrammatic calculation:
For a tolerance of $\tau=10^{-3}$, the ranks of the Matsubara $\kclass_2^t$ component are no larger than 20.
The variation of the rank with $\beta$ and $u$ does therefore not bare much significance.
For smaller tolerances, the rank of $\kclass_2^t$ remains much lower than that of the core vertex.
Analogous observations hold for the Keldysh $\kclass_2^t$ component with a target tolerance of $\tau=10^{-2}$. Like the Keldysh core vertex, its rank
increases slowly with $\beta$ and decreases for strong coupling $u$, but the changes are small
compared to the core vertex.

\section{Compression for different channels, flavors, Keldysh components}
\label{app:channelflavorkeldysh}
In this section we investigate the compressibility of core and full vertices for different flavors ($\upup,\updown$),
frequency channels ($r=a,p,t$) and Keldysh components $\boldsymbol{k}=(k_1k_2k_3k_4)$.

\Fig{fig:ranksvsflavorchannel} shows the QTT ranks of the Matsubara and Keldysh full and core vertices
for different channels and flavors, at $\beta=2000$ and $u=0.5$.
The tolerances are $\tau=10^{-3}$ and $\tau=10^{-2}$ for Matsubara and Keldysh vertices, respectively.
We observe that the $p$-channel exhibits the highest ranks throughout, which is why we used this
frequency parametrization in the main text.
The QTT ranks of the Matsubara vertices shown in \Fig{fig:ranksvsflavorchannel}(c) barely differ between the two flavors.
By contrast, $\gamcore^{2121,\upup}$ has a significantly higher rank than $\gamcore^{2121,\updown}$ ($\chi=198$ vs.\ $\chi=154$).

The rank of $\gamcore^{\upup}$ and $\Gamma^{\upup}$ depending on the Keldysh component is shown in \Fig{fig:RanksvsiK}.
Only components that are not related by crossing or complex conjugation symmetry \cite{Jakobs2010} are considered.
We show data for $\beta=2000$, $u=0.5$ and $\tau=10^{-3}$.
The $(2121)$ component of $\gamcore$ is found to have the highest rank.
We therefore selected $\gamcore^{2121,\upup}$ for our analysis of rank saturation and parameter dependence in
Figs.\ \ref{fig:KeldyshVertex_tolvsrank} and \ref{fig:KeldyshParamsVSRank}.

\section{Frequency conventions}
\label{app:frequency_conventions}
We use the following parametrizations for the $t$ (particle-hole), $p$ (particle-particle) and $a$ (transverse particle-hole) channels:
\begin{equation}
    \bomega=\begin{cases}
        (-\nu_r,\omega_r+\nu_r,-\omega_r-\nu'_r,\nu'_r)\quad&\text{for  $r=t$ (ph)},\\
        (-\nu_r,\omega_r-\nu'_r,-\omega_r+\nu_r,\nu'_r)\quad&\text{for  $r=p$ (pp)},\\
        (-\nu_r,\nu'_r,-\omega_r-\nu'_r,\omega_r+\nu_r)\quad&\text{for $r=a$ ($\overline{\text{ph}}$)}.
    \end{cases}
\end{equation}
These are the same as in Ref.\ \onlinecite{sIE2024}, up to a global minus sign.
Spin and, if present, Keldysh indices of the vertex $\Gamma^{\boldsymbol{k},\sigma_1\sigma_2\sigma_3\sigma_4}(\bomega)$ are ordered according to the underlying impurity Green's function $G^{\boldsymbol{k}}_{\mathrm{con}}[d_{\sigma_1}d_{\sigma_2}^{\dag}d_{\sigma_3}d_{\sigma_4}^{\dag}](\bomega)$. Finally, the evaluation of a 2p correlator $G$ at a frequency $\nu$ is defined as $G(\nu,-\nu)$ in our convention. This is relevant for evaluating the self-energy $\Sigma$ in \Eq{eq:Ysymbol}, because computing the self-energy comes down to evaluating 2p correlators according to the asymmetric estimators (see Eq.\ (27) in Ref.\ \cite{sIE2024}) we employed.

\section{Evaluating the Keldysh vertex on nonlinear grids}
\label{app:nonlin_V_KF}
As mentioned in \Sec{subsec:PSFtoCorr}, the PSFs consist of spectral peaks residing on a logarithmic energy grid. This raises the question whether the vertices could also be computed on logarithmic (or arbitrary nonlinear) grids. Since the Matsubara vertex is inherently defined on an equidistant grid, we consider this question to be relevant only for the Keldysh vertex. While the spectral representation and symmetric estimators yield the Keldysh vertex at arbitrary frequency points, evaluating the vertex on a nonlinear grid turns out to be much more expensive. This section explains why that is the case.

Consider a grid $\onedgrid\subset\mathbb{R}$, where the frequencies $\omega\in\onedgrid$ may be spaced, e.g., logarithmically. Suppose we want to evaluate the Keldysh vertex on points
$\bomega\in {\onedgrid}^3\subset\mathbb{R}^3$. As detailed in Secs.\ \ref{subsec:keldysh_formalism} and \ref{subsec:sIE}, this entails the evaluation of full correlators $G^{\boldsymbol{k}}(\bomega)$ with $\bomega\in {\onedgrid}^3$, which in turn requires $G^{[\lambda]}_p(\bomega_p)$ (see \Eq{eq:KeldyshContraction}). Crucially, the broadened kernels $k_b^{[\lambda,i]}$ are evaluated at the transformed frequency $\omint{i}$ in \Eq{eq:KeldyshContraction}.
Let us denote this linear transformation $\bomega\mapsto\left(\bomintthree\right)$ for 4p correlators by $T_p$. The broadened kernels $k_b^{[\lambda,i]}$ must be evaluated on points in the image of the chosen grid ${\onedgrid}^3$ under $T_p$, i.e., $T_p({\onedgrid}^3)\ni \bomega_p$. In order to save computation time by precomputing the kernels, one can embed the image grid in a Cartesian product of one-dimensional grids ${\onedgrid}_{p,1},{\onedgrid}_{p,2},{\onedgrid}_{p,3}\subset \mathbb{R}$, i.e., $T_p({\onedgrid}^3)\subset {\onedgrid}_{p,1}\times{\onedgrid}_{p,2}\times{\onedgrid}_{p,3}$. It is the grids ${\onedgrid}_{p,i}$ on which the kernels are broadened and, if desired, SVD compressed as explained in \Sec{subsec:keldysh_implementation_details}.
In the case of an \textit{equidistant} grid, the ${\onedgrid}_{p,i}$ have at most $3\cdot|\onedgrid|=\order{|\onedgrid|}$ points.

By constrast, if $\onedgrid$ is nonlinear, one has $|{\onedgrid}_{p,i}|=\order{|\onedgrid|^3}$ in general. This means that, for large grid sizes, precomputing and storing the broadened kernels $k_b^{[\lambda,i]}$ on ${\onedgrid}_p$ is no longer possible. The same holds for the SVD compression of the kernels. This issue leaves two alternatives for evaluating the Keldysh vertex $\Gamma^{\boldsymbol{k}}$ on nonlinear grids:

(i) For each individual point $\bomega\in{\onedgrid}^3$, compute the broadened kernels upon evaluating $\Gamma^{\boldsymbol{k}}(\bomega)$. This is not implemented in our code \cite{TCI4Keldysh}, but we expect it to be prohibitively expensive even if caching of previously computed kernel values were introduced.

(ii) Precompute the kernels on a very fine, equidistant grid and evaluate the vertex by interpolation from linear grids. This approach is implemented in our code \cite{TCI4Keldysh} and has been employed in Ref.\ \onlinecite{Ritz2025} when investigating the fulfillment of diagrammatic identities by mpNRG data.

In conclusion, we do not see an affordable method of evaluating the Keldysh vertex 'directly' on a large, nonlinear grid. However, interpolation from a linear grid provides a viable alternative: Using QTCI, one can determine at which grid size the vertex is resolved to the desired accuracy and then use the resulting quantics tensor train for interpolation.

Finally, we report a preliminary investigation of the QTCI-compressibility of Keldysh vertex components stored on logarithmic grids: We compressed different Keldysh components of the $t$-channel Keldysh core vertex at $\beta=2000,\,u=0.50$ computed on a $243\times243\times243$ logarithmic grid. The TCI tolerance was $\tcitoltxt10^{-2}$. The grid spans a large frequency range from $-3.18$ to $3.18$. The vertex was interpolated trilinearly from a quantics grid with $R=15$ bits, see Ref.\ \onlinecite{Ritz2025}.
The results are summarized in Tab.\ \ref{tab:nonlin_bonddims}, showing large QTT ranks $\chi_{\mathrm{log}}$ compared to the QTT ranks on a linear grid, $\chi_{\mathrm{lin}}$. This indicates that a QTCI compression of Keldysh vertex data on a logarithmic grid is not effective.

\begin{table}[h]
    \centering
    \setlength{\tabcolsep}{4pt}
    \begin{tabular}{c|ccccc}
        $\boldsymbol{k}$ & $1111$ & $2111$ & $2121$ & $2112$ & $1222$ \\
        \midrule
        $\chi_{\mathrm{log}}$ & 258 & 475 & 718 & 379 & 445 \\
        $\chi_{\mathrm{lin}}$ & 129 & 150 & 177 & 127 & 101
    \end{tabular}
    \caption{
    QTT ranks of different components of the $t$-channel Keldysh core vertex $\Gamma_{\mathrm{core}}^{\upup}$ computed on a logarithmic grid, and, for comparison, on an $R=15$ linear quantics grid. The tolerance is $\tau=10^{-2}$.}
    \label{tab:nonlin_bonddims}
\end{table}
\FloatBarrier
\bibliography{references}

\end{document}